\documentclass[smallextended]{svjour3}       
\smartqed  

\RequirePackage[OT1]{fontenc}
\usepackage{amsmath}
\usepackage{algorithm}
\usepackage{algorithmic}
\usepackage{parselines}
\usepackage{epsfig}
\usepackage[nottoc]{tocbibind}
\usepackage[colorlinks,citecolor=blue,urlcolor=blue,filecolor=blue,backref=page]{hyperref}
\usepackage{graphicx}
\usepackage{graphics}
\usepackage{subfigure}
\usepackage{verbatim}
\usepackage{float}
\usepackage{mathtools}
\usepackage{amssymb}
\usepackage{enumerate}
\usepackage{natbib}
\usepackage{url}

 \usepackage{mathptmx}      
\journalname{}

\begin{document}

\title{Some variations of EM algorithms for Marshall-Olkin bivariate Pareto distribution with location and scale} 


\titlerunning{EM-BVPA}        
\author{ Arabin Kumar Dey \and Biplab Paul
}

           

\institute{A. K. Dey \at
              Department of Mathematics, \\
              IIT Guwahati,\\
              Guwahati, India\\
              Assam\\
              Tel.: +91361-258-4620\\
              \email{arabin@iitg.ac.in}\\
            \and
            B. Paul \at
              Department of Mathematics,\\ 
             IIT Guwahati,\\   
             Guwahati, India\\
             \email{biplab.paul@iitg.ac.in}            
}






\maketitle

\begin{abstract}

 Recently Asimit et. al (\cite{AsimitFurmanVernic:2016}) used an EM algorithm to estimate Marshall-Olkin bivariate Pareto distribution.  The distribution has seven parameters.  We describe few alternative approaches of EM algorithm.  A numerical simulation is performed to verify the performance of different proposed algorithms.   A real-life data analysis is also shown for illustrative purposes.

\end{abstract}


\section{Introduction}
\label{intro}

 Bivariate Pareto has several forms.  In this paper we study Marshal-Olkin formulation (\cite{MarshallOlkin:1967}) of this distribution which includes both location and scale parameters.  In a very recent paper, Asimit et al. (\cite{AsimitFurmanVernic:2010}) used EM algorithm to estimate the parameters of this distribution.  We adapt few more variations of the same.  We observe that some of our proposed algorithms either works equally well or outperforms their algorithm.  We also present some real life data analysis which is absent in their paper.        

 In this paper our focus is on \textbf{singular} bivariate Pareto distribution whose both the marginals have Pareto type-II distribution.  We can obtain bivariate Pareto distribution considering peak over threshold method in a bivariate data.  The distribution has wide application in modeling data related to finance, insurance, environmental sciences and internet network.  Any analysis based on this distribution requires efficient techniques for estimating parameters of the distribution.  We consider a more generalized set up including location and scale parameters in formation of Marshal-Olkin bivariate Pareto distribution.  This dependence structure can be described by well-known Marshall-Olkin copula too  [\cite{Nelsen:2007}, \cite{MarshallOlkin:1967}, \cite{Yeh:2000}, \cite{Yeh:2004}].  We propose few innovative ways to implement EM algorithm.

 A random variable X is said to have Pareto of second kind, i.e. $X \sim Pa(II)(\mu, \sigma, \alpha)$ if it has the survival function  
$$ \bar{F}_{X}(x ; \mu, \sigma, \alpha) = P(X > x) = (1 + \frac{x - \mu}{\sigma})^{-\alpha} $$
and the probability density function (pdf) $$ f(x ; \mu, \sigma, \alpha) = \frac{\alpha}{\sigma}(1 + \frac{x - \mu}{\sigma})^{-\alpha - 1} $$
with $x > \mu \in \mathcal{R}$, $\sigma > 0$ and $\alpha > 0$.

 \cite{KunduGupta:2009}, \cite{KunduGupta:2010}, \cite{KunduDey:2009} used EM algorithm for estimating parameters of different bivariate distributions, e.g. bivariate generalized exponential, bivariate Weibull etc.  \cite{DeyKundu:2012} performed discrimination between bivariate Weibull and bivariate generalized exponential distributions where they used parameter estimation through EM algorithm. \cite{SarhanBalakrishnan:2007} considered estimation issues in Sarhan and Balakrishnan bivariate distribution with extra scale parameter in their model.  Many works have been done on multivariate Pareto distribution too [\cite{Hanagal:1996}, \cite{Yeh:2000}, \cite{Yeh:2004}, \cite{AsimitFurmanVernic:2010}].  Statistical Inference of multivariate Pareto distribution through EM algorithm is attempted by \cite{AsimitFurmanVernic:2016}.  This paper also deals with seven parameters including location and scale as its parameters.  We handle the same problem in a slightly different manner.  

 We arrange the paper in the following way.  In section 2 we keep the Marshall-Olkin bivariate Pareto formulation and some of its properties.  In section 3, we describe our proposed EM algorithms.  Some simulation results show the performance of the algorithm in section 4.  In section 5 we show the data analysis.  Finally we conclude the paper in section 6.

\section{Formulation of Marshal-Olkin bivariate Pareto}

  Let $ U_{0}$, $U_{1}$ and $U_{2}$ be three independent univariate type-II Pareto distributions $Pa(II)(0, 1, \alpha_{0})$, $Pa(II)(\mu_{1}, \sigma_{1}, \alpha_{1})$ and $Pa(II)(\mu_{2}, \sigma_{2}, \alpha_{2})$.  

  We define $X_{1} = \min\{ \sigma_{1} U_{0} + \mu_{1}, U_{1} \}$ and $X_{2} = \min\{ \sigma_{2} U_{0} + \mu_{2}, U_{2} \}$.  We can show that $(X_{1}, X_{2})$ jointly follow bivariate Pareto distribution of second kind, we call it as $BVPA(\mu_{1}, \mu_{2}, \sigma_{1}, \sigma_{2}, \alpha_{0}, \alpha_{1}, \alpha_{2})$

The joint distribution can be given by 
\begin{eqnarray*}
 f(x_{1},x_{2}) = \begin{cases}  f_{1}(x_{1}, x_{2})  & \text{if $\frac{x_{1} - \mu_{1}}{\sigma_{1}} < \frac{x_{2} - \mu_{2}}{\sigma_{2}}$}\\
 f_{2}(x_{1}, x_{2})  & \text{if $\frac{x_{1} - \mu_{1}}{\sigma_{1}} > \frac{x_{2} - \mu_{2}}{\sigma_{2}}$}\\
 f_{0}(x) & \text{if $\frac{x_{1} - \mu_{1}}{\sigma_{1}}= \frac{x_{2} - \mu_{2}}{\sigma_{2}} = x $}
\end{cases}
\end{eqnarray*}
where 
\begin{eqnarray*} f_{1}(x_{1}, x_{2}) & = & \frac{\alpha_{1}}{\sigma_{1}\sigma_{2}}(\alpha_{0}+\alpha_{2})(1+\frac{x_{2} - \mu_{2}}{\sigma_{2}})^{-(\alpha_{0} + \alpha_{2} + 1)}(1 + \frac{x_{1} - \mu_{1}}{\sigma_{1}})^{-(\alpha_{1} + 1)}\\
 f_{2}(x_{1}, x_{2}) & = & \frac{\alpha_{2}}{\sigma_{1}\sigma_{2}}(\alpha_{0} + \alpha_{1})(1 + \frac{x_{1} - \mu_{1}}{\sigma_{1}})^{-(\alpha_{0} + \alpha_{1} + 1)}(1 + \frac{x_{2} - \mu_{2}}{\sigma_{2}})^{-(\alpha_{2} + 1)}\\
 f_{0}(x) & = & \alpha_{0}(1 + x)^{-(\alpha_{1} + \alpha_{2} + \alpha_{0} + 1)} \end{eqnarray*}

It can be shown that 
\begin{enumerate}

\item The distribution of $X_{j}$ is $Pa(II)(\mu_{j}, \sigma_{j}, \alpha_{0j})$,  $\alpha_{0j} = \alpha_{0} + \alpha_{j}$. 

\item Distribution of the minimum is also Pareto distribution when $U_{0}$, $U_{1}$ and $U_{2}$ has same location and scale parameter.

\item Maximum Likelihood estimate of $\mu, \sigma, \alpha$, the parameters of univariate Pareto can be given based on the data set $Y_{1}, Y_{2}, \cdots, Y_{n}$ as : 
$\mu = \min\{ Y_{1}, Y_{2}, \cdots, Y_{n} \}$, whereas estimates of $\alpha$ and $\sigma$ can be obtained by solving the fixed point iterations :  \begin{eqnarray*} \sigma = \frac{\alpha + 1}{n} \sum_{i = 1}^{n} \frac{(x_{i} - \mu)}{(1 + \frac{x_{i} - \mu}{\sigma})} \end{eqnarray*} where, \begin{eqnarray*}  \alpha = \frac{n}{\sum_{i = 1}^{n} \ln(1 + \frac{x_{i} - \mu}{\sigma})} \end{eqnarray*}
\end{enumerate}

\begin{figure}[H]
 \begin{center}
  \subfigure[$\xi_{1}$]{\includegraphics[angle = -90,width = 0.45\textwidth]{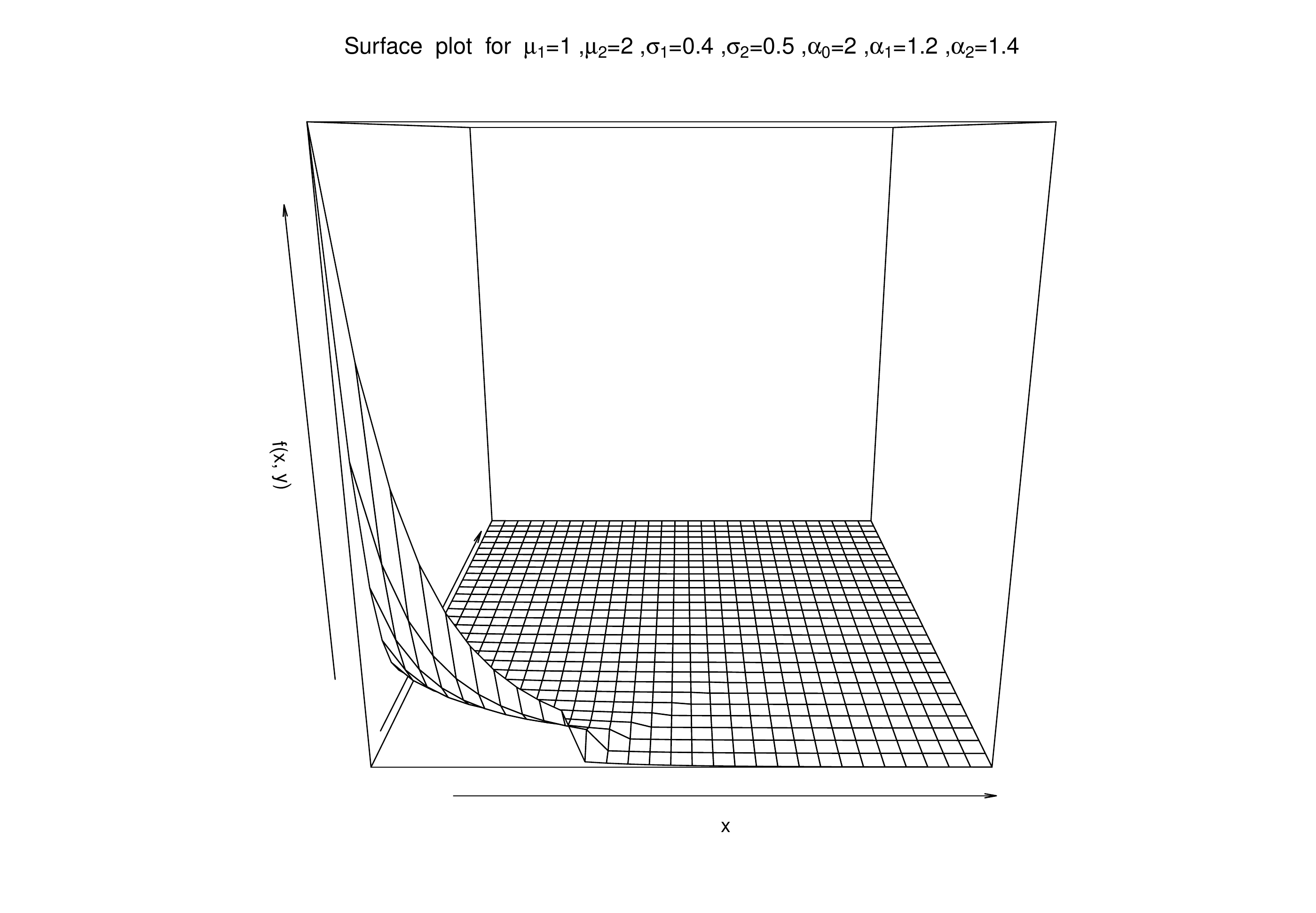}}
  \subfigure[$\xi_{2}$]{\includegraphics[angle = -90, width = 0.45\textwidth]{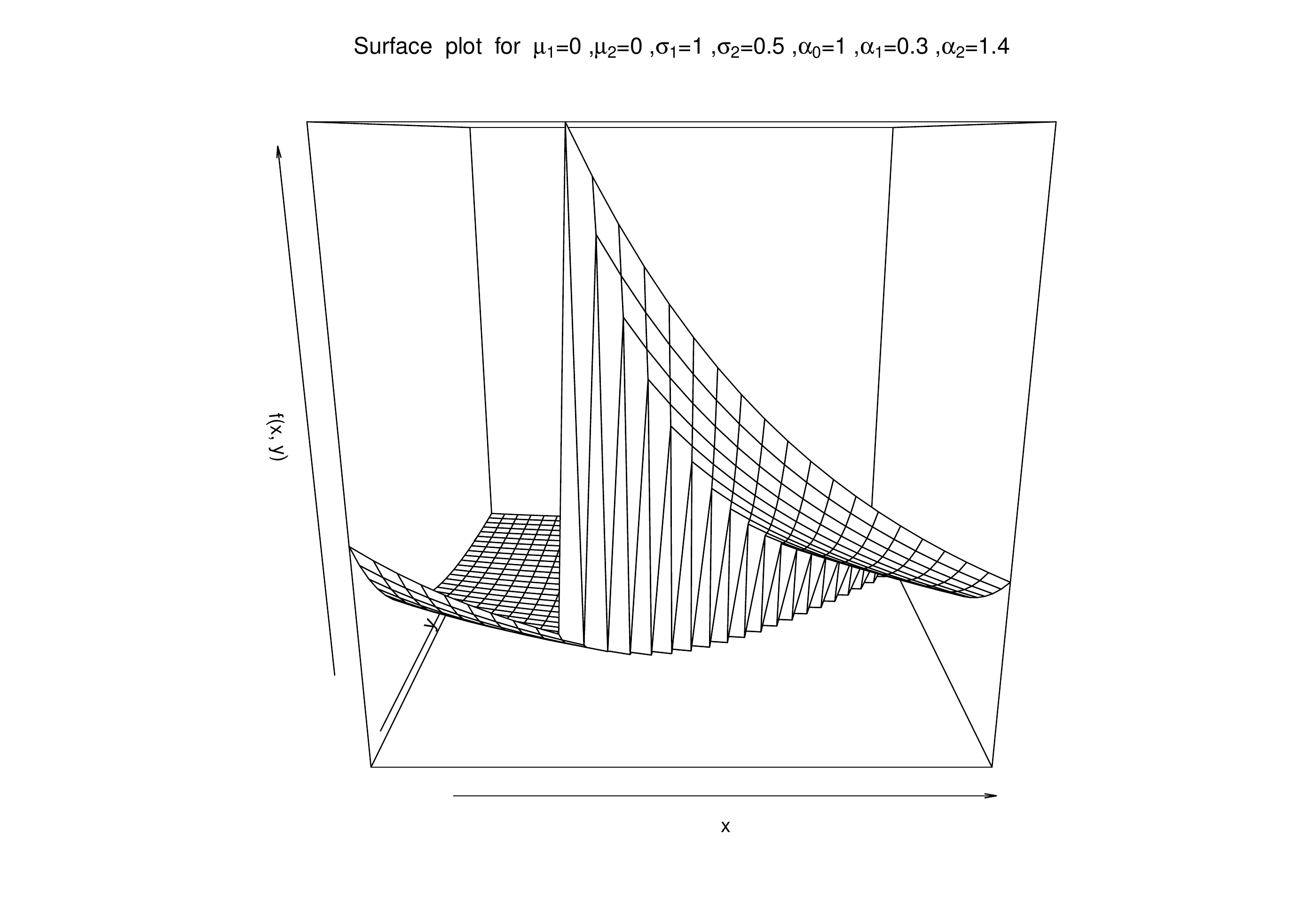}}\\
  \subfigure[$\xi_{3}$]{\includegraphics[angle = -90, width = 0.45\textwidth]{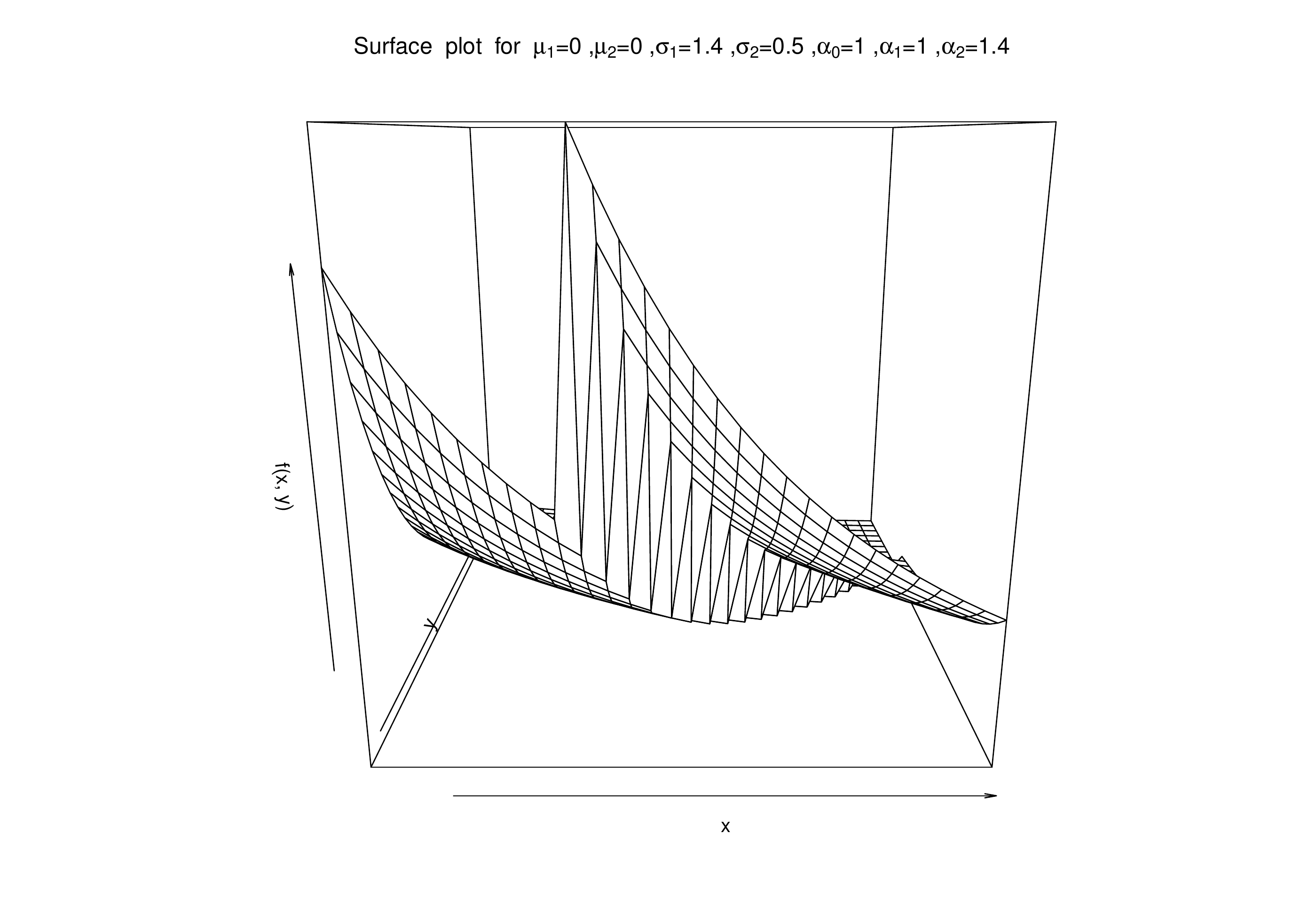}}
  \subfigure[$\xi_{4}$]{\includegraphics[angle = -90, width = 0.45\textwidth]{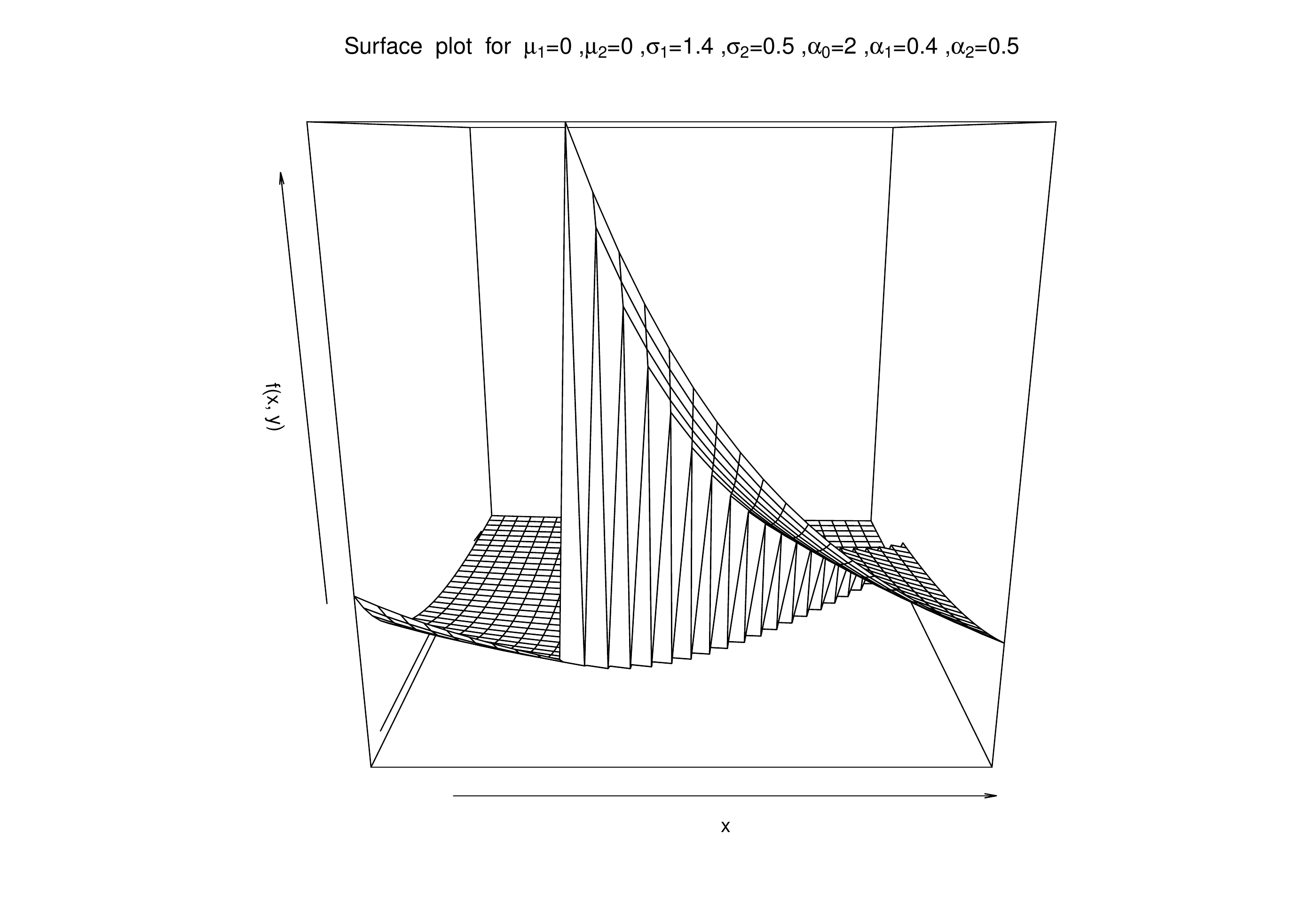}}\\
\caption{Surface plots for pdf of BVPA \label{fig1}}
\end{center}
\end{figure}

\begin{figure}[H]
 \begin{center}
  \subfigure[$\xi_{1}$]{\includegraphics[angle = -90,width = 0.45\textwidth]{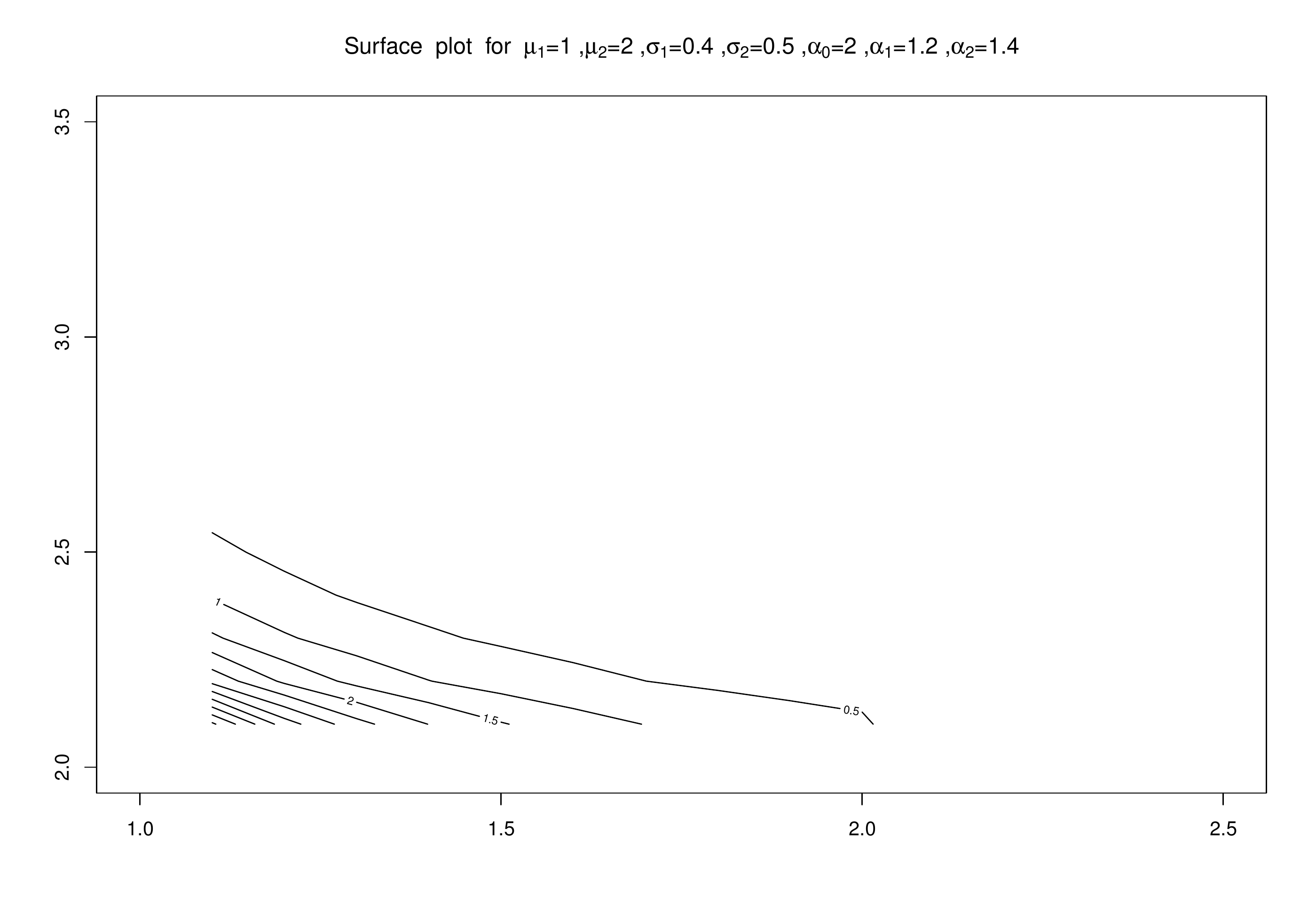}}
  \subfigure[$\xi_{2}$]{\includegraphics[angle = -90, width = 0.45\textwidth]{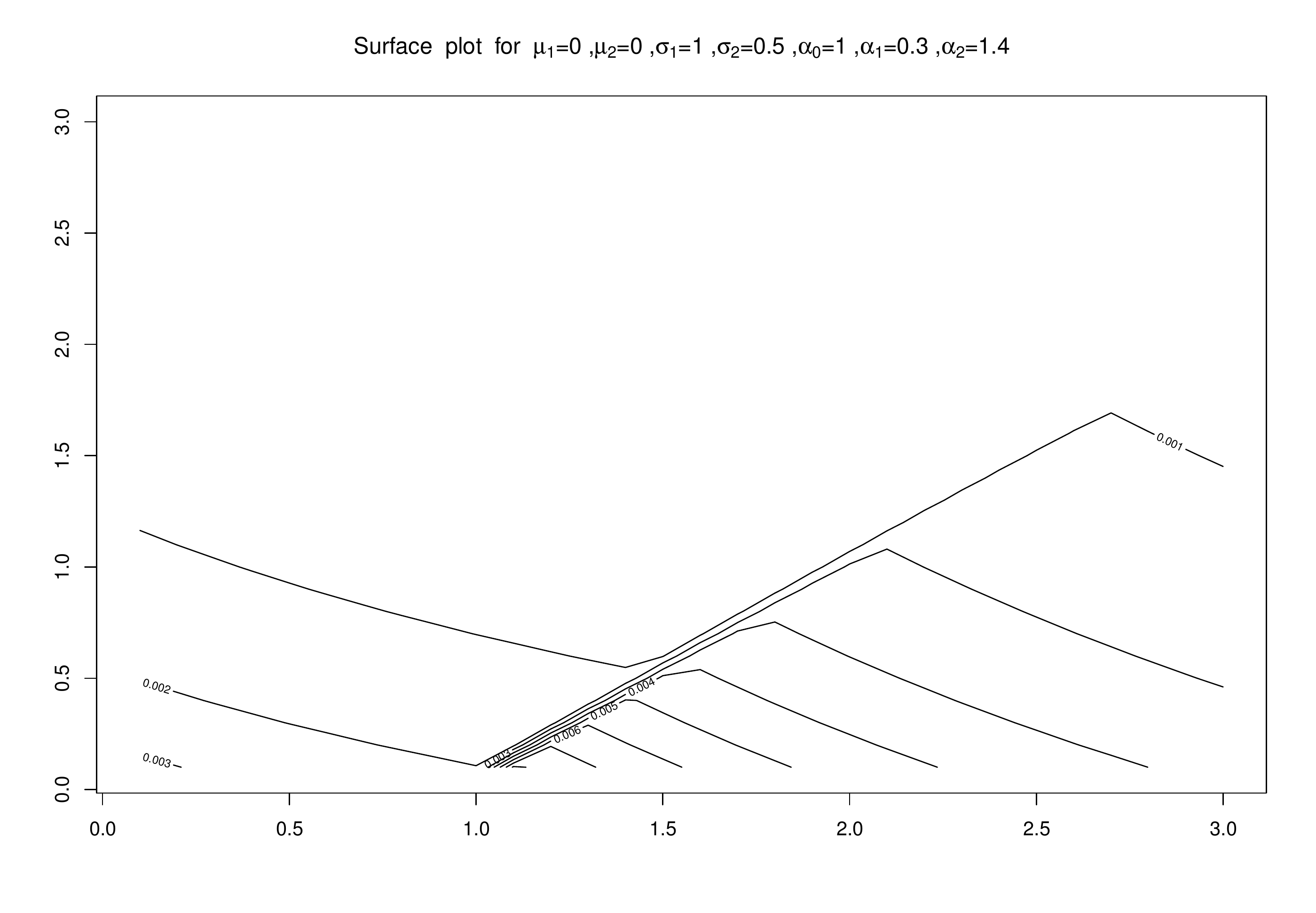}}\\
  \subfigure[$\xi_{3}$]{\includegraphics[angle = -90, width = 0.45\textwidth]{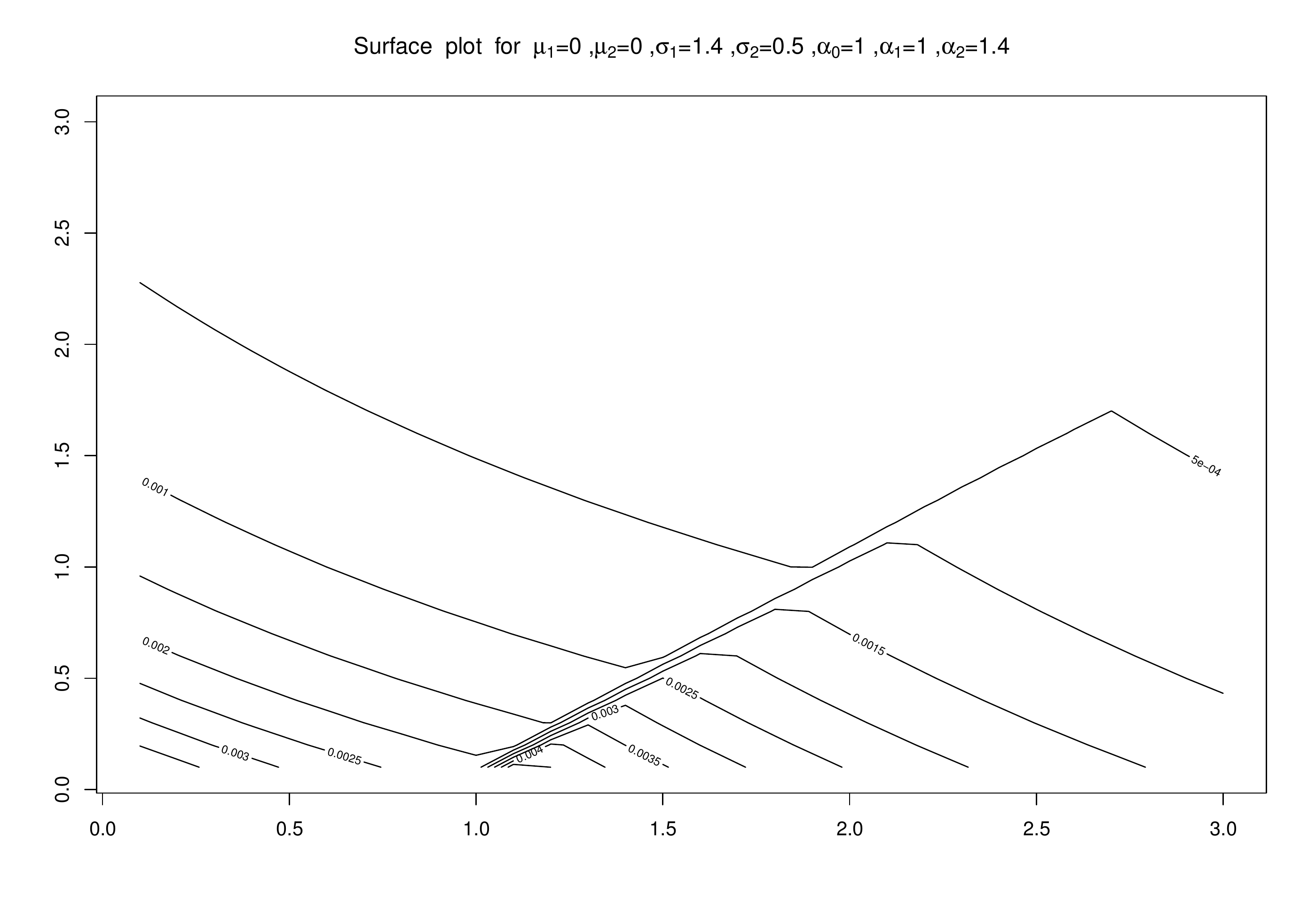}}
  \subfigure[$\xi_{4}$]{\includegraphics[angle = -90, width = 0.45\textwidth]{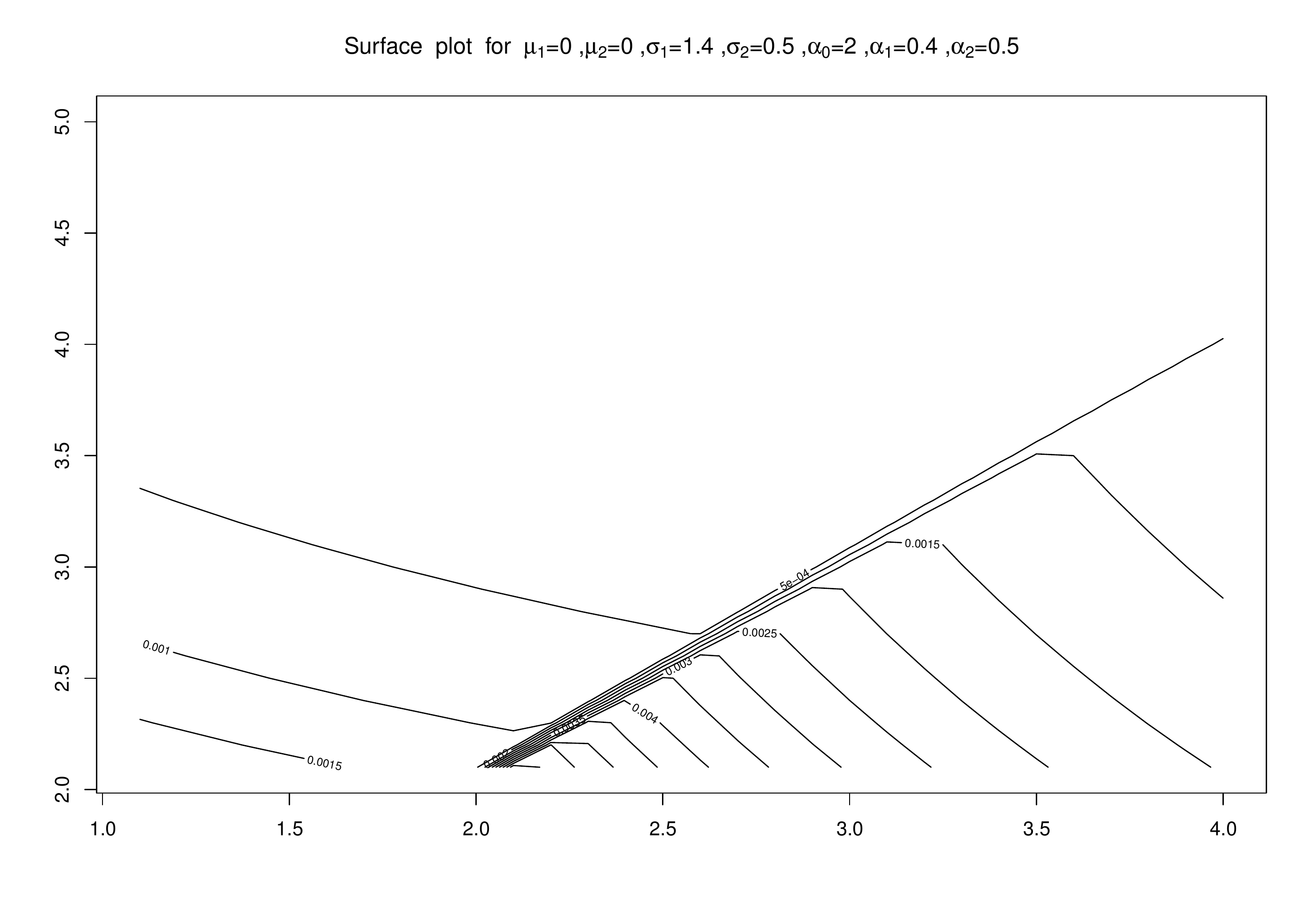}}\\
\caption{Contour plots for pdf of BVPA \label{fig2}}
\end{center}
\end{figure}

 Surface and contour plots of the absolutely continuous part of the pdf are shown in Figure-\ref{fig1} and Figure-\ref{fig2} respectively. The following four different sets of parameters provide four subfigures in each figure.
$ \xi_{1} : \mu_1 = 0, \mu_2 = 0, \sigma_1 = 1, \sigma_2 = 0.5, \alpha_0 = 1, \alpha_1 = 0.3, \alpha_2 = 1.4;$
$ \xi_{2} : \mu_1 = 1, \mu_2 = 2, \sigma_1 = 0.4, \sigma_2 = 0.5, \alpha_0 = 2, \alpha_1 = 1.2, \alpha_2 = 1.4;$
$ \xi_{3} :\mu_1 = 0, \mu_2 = 0, \sigma_1 = 1.4, \sigma_2 = 0.5, \alpha_0 = 1, \alpha_1 = 1, \alpha_2 = 1.4;$
$ \xi_{4} : \mu_1 = 0, \mu_2 = 0, \sigma_1 = 1.4, \sigma_2 = 0.5, \alpha_0 = 2, \alpha_1 = 0.4, \alpha_2 = 0.5.$

\section{EM-algorithm}

 Let us assume $\mu_{1}$, $\mu_{2}$, $\sigma_{1}$ and $\sigma_{2}$ are known.  Now we divide our data into three parts :-

$ I_0 = \{ (x_{1i},  x_{2i}) :\frac{x_{1i} - \mu_{1}}{\sigma_{1}} = \frac{x_{2i} - \mu_{2}}{\sigma_{2}} \}$ ,$ I_1 = \{ (x_{1i} , x_{2i} ) : \frac{x_{1i} - \mu_{1}}{\sigma_{1}} < \frac{x_{2i} - \mu_{2}}{\sigma_{2}} \}$ , $ I_2 = \{ (x_{1i} , x_{2i} ) : \frac{x_{1i} - \mu_{1}}{\sigma_{1}} > \frac{x_{2i} - \mu_{2}}{\sigma_{2}} \} $

Usual Likelihood function for the parameters of the BVPA can be written as 
\begin{eqnarray*}
&& L(\mu_{1}, \mu_{2}, \sigma_{1}, \sigma_{2}, \alpha_{0}, \alpha_{1}, \alpha_{2})\\ & = & n_{1}\ln\alpha_{1} + n_{1}\ln(\alpha_{0} + \alpha_{2}) - n_{1}\ln\sigma_{1} - n_{1}\ln\sigma_{2}\\ & - & (\alpha_{0} + \alpha_{2} + 1)\sum_{i \in I_{1}}^{n} \ln(1 + \frac{x_{2i} - \mu_{2}}{\sigma_{2}}) -  (\alpha_{1} + 1)\sum_{i \in I_{1}}^{} \ln(1 + \frac{x_{1i} - \mu_{1}}{\sigma_{1}})\\ & - & n_{2}\ln\sigma_{1}  - n_{2}\ln\sigma_{2} +  n_{2}\ln\alpha_{2} + n_{2}\ln(\alpha_{0} + \alpha_{1})\\ & - & (\alpha_{0} + \alpha_{1} + 1)\sum_{i \in I_{2}}^{n} \ln (1 + \frac{x_{1i} - \mu_{1}}{\sigma_{1}}) - (\alpha_{2} + 1)\sum_{i \in I_{2}}^{n} \ln(1 + \frac{x_{2i} - \mu_{2}}{\sigma_{2}})\\ & + & n_{0}\ln\alpha_{0} - n_{0}\ln\sigma_{1} - (\alpha_{0} + \alpha_{1} + \alpha_{2} + 1)\sum_{i \in I_{0}}^{n} \ln(1 + \frac{x_{1i} - \mu_{1}}{\sigma_{1}})
\end{eqnarray*}  

 Direct maximum likelihood estimates of the parameters based on $(X_{1i}, X_{2i}) : i = 1(1)n$ may not be simple.  We implement EM algorithm first given some $\mu_{1}$, $\mu_{2}$, $\sigma_{1}$ and $\sigma_{2}$.  It requires identification of some missing structure within the problem.      

 We do not know $X_1$ is $\sigma_{1} U_{0} + \mu_{1}$ or $U_1$ and We do not know $X_2$ is $\sigma_{2} U_{0} + \mu_{2}$ or $U_2$.  So we introduce two new random variables ($\Delta_1 ; \Delta_2 $) as
$$ \Delta_1 = \begin{cases}  0  & \text{if $X_1 = \sigma_{1} U_{0} + \mu_{1}$  }\\ 1  & \text{if $X_1 = U_1$  }\\  \end{cases} $$
            and
$$ \Delta_2 = \begin{cases}  0  & \text{if $X_2 = \sigma_{2} U_{0} + \mu_{2}$  }\\ 2  & \text{if $X_2 = U_2$  }\\  \end{cases} $$

The $\Delta_1$ and $\Delta_2$ are the missing values of the E-M algorithm. To calculate the E-step we need the conditional distribution of $\Delta_1$ and $\Delta_2$.

Using the definition of $X_{1}$, $X_{2}$, $\Delta_1$ and $\Delta_2$ we have :

$\vartriangleright \text{For group } I_0, $ both $\Delta_1$ and $\Delta_2$ \text{are known,}\\
$$\Delta_1 = \Delta_2 = 0 $$ 

$\vartriangleright \text{For group }  I_1, $ $\Delta_1$ is known, $\Delta_2$ is unknown,
$$\Delta_1 = 1, \Delta_2 = 0 ~\text{or } 2 $$
       
Therefore we need to find out $u_1 = P(\Delta_2 = 0| I_1 ) \text{and } u_2 = P (\Delta_2 = 2|I_1 )$\\
$\vartriangleright \text{For group } I_2, $ $\Delta_2$ is known, $\Delta_1$ is unknown,
$$ \Delta_1 = 0 ~\text{or }  1, \Delta_2 = 2$$
Moreover we need $w_1 = P (\Delta_1 = 0 | I_2 )$ and $w_2 = P (\Delta_1 = 1 | I_2 )$

  Since, each posterior probability corresponds to one of the ordering from Table-\ref{ordering}, we calculate $u_1, u_2, w_1, w_2$ using the probability of appropriate ordering, where $U^{*}_{1} = \frac{(U_{1} - \mu_{1})}{\sigma_{1}}$, $U^{*}_{2} = \frac{(U_{2} - \mu_{2})}{\sigma_{2}}$.

\begin{table}[H]
\begin{center}
\begin{tabular}{|c|c|r|}
	\hline
Ordering &  $(X_1, X_2)$   & Group  \\
	\hline
$U_0 < U^{*}_1 < U^{*}_2$  & $(U_0,U_0)$ & $I_0$\\
$U_0 < U^{*}_2 < U^{*}_1$  & $(U_0,U_0)$ & $I_0$\\
$U^{*}_1 < U_0 < U^{*}_2$  & $(U^{*}_1,U_0)$ & $I_1$\\
$U^{*}_1 < U^{*}_2 < U_0$  & $(U^{*}_1,U^{*}_2)$ & $I_1$\\
$U^{*}_2 < U_0 < U^{*}_1$  & $(U_0,U^{*}_2)$ & $I_2$\\
$U^{*}_2 < U^{*}_1 < U_0$  & $(U^{*}_1,U^{*}_2)$ & $I_2$\\
	\hline
\end{tabular}
\caption{Groups and corresponding orderings of hidden random variables $U_0$, $U^{*}_{1}$ and $U^{*}_{2}$}  
\label{ordering}
\end{center}
\end{table}

We have the following expressions for $u_1, u_2, w_1, w_2$ 
$$u_1=\frac{P(U^{*}_1 < U_0 < U^{*}_2)}{P(U^{*}_1 < U_0 < U^{*}_2) + P(U^{*}_1 < U^{*}_2 < U_0)}$$
$$u_2=\frac{P(U^{*}_1 < U^{*}_2 < U_0)}{P(U^{*}_1 < U_0 < U^{*}_2) + P(U^{*}_1 < U^{*}_2 < U_0)}$$
$$w_1=\frac{P(U^{*}_2 < U_0 < U^{*}_1)}{P(U^{*}_2 < U_0 < U^{*}_1) + P(U^{*}_2 < U^{*}_1 < U_0)}$$
$$w_2=\frac{P(U^{*}_2 < U^{*}_1 < U_0)}{P(U^{*}_2 < U_0 < U^{*}_1) + P(U^{*}_2 < U^{*}_1 < U_0)}$$
Now we have 
\begin{align*}
P(U^{*}_1 < U_0 < U^{*}_2) &=\int_0^\infty [1-(1 + x)^{-\alpha_1}] \alpha_0 (1 + x)^{-(\alpha_0 + 1)}(1 + x)^{-(\alpha_2)} \\
                   &=\alpha_0 \int_0^\infty (1 + x)^{-(\alpha_0 + \alpha_2 + 1)} - (1 + x)^{-(\alpha_0 + \alpha_1 + \alpha_2 +1)} \\
                   &=\frac{\alpha_0 \alpha_1}{(\alpha_0 + \alpha_2 + 1)(\alpha_0 + \alpha_1 + \alpha_2 + 1 )}\\
\end{align*}
Using this above result we evaluate the other probabilites to get values of $u_1, u_2, w_1, w_2$ as :
$u_1 = \frac{\alpha_0}{\alpha_0 + \alpha_2} \text{ and } u_2 = \frac{\alpha_2}{\alpha_0 + \alpha_2}$
$w_1 = \frac{\alpha_0}{\alpha_0 + \alpha_1} \text{ and } w_2 = \frac{\alpha_1}{\alpha_0 + \alpha_1}$

\subsection{Pseudo-likelihood expression}

We define $n_0, n_1, n_2 $ as :
$n_0 = |I_0|, n_1 = |I_1|, n_2 = |I_2| $
where $|I_j|$ for $j = 0, 1, 2$ denotes the number of elements in the set $I_j$. 
Now the pseudo log-likelihood can be written down as
\begin{eqnarray}
 Q & = & L(\alpha_{0}, \alpha_{1}, \alpha_{2})\nonumber\\ & = & -\alpha_0 (\sum\limits_{i\in I_0} \ln (1 + x_i) + \sum\limits_{i\in I_2} \ln (1 + x_{1i}) + \sum\limits_{i\in I_1} \ln (1 + x_{2i})) \nonumber\\ & + & (n_0 + u_1n_1 + w_1n_2) \ln \alpha_0  - \alpha_1(\sum\limits_{i\in I_0} \ln (1 + x_i) + \sum\limits_{i\in I_1 \cup I_2} \ln (1 + x_{1i})) \nonumber\\ & + & (n_1 + w_2n_2)\ln \alpha_1  - \alpha_2(\sum\limits_{i\in I_0} \ln (1 + x_i) + \sum\limits_{i\in I_1\cup I_2} \ln (1 + x_{2i})) \nonumber\\ & + & (n_2 + u_2n_1)\ln \alpha_2 
\label{likelihood}
\end{eqnarray}

  Therefore M-step involves maximizing (\ref{likelihood}) with respect to $\alpha_{0}$, $\alpha_{1}$, $\alpha_{2}$ at 

\begin{equation} \hat{\alpha}^{(t + 1)}_{0} = \frac{n_{0} + u^{(t)}_{1} n_{1} + w^{(t)}_{1} n_{2}}{\sum_{i \in I_{0}}^{} \ln(1 + x_{i}) + \sum_{i \in I_{2}}^{} \ln(1 + x_{1i}) + \sum_{i \in I_{1}}^{} \ln (1 + x_{2i})}. \label{alph0}\end{equation}
\begin{equation} \hat{\alpha}^{(t + 1)}_{1} = \frac{(n_{1} + w^{(t)}_{2}n_{2})}{\sum_{i \in I_{0}}^{} \ln(1 + x_{i})  + \sum_{i \in (I_{1} \cup I_{2})}^{} \ln(1 + x_{1i}) } \label{alph1}\end{equation}
\begin{equation} \hat{\alpha}^{(t + 1)}_{2} = \frac{n_{2} + u^{(t)}_{2} n_{1}}{(\sum_{i \in I_{0}}^{} \ln(1 + x_{i}) +  \sum_{i \in (I_{1} \cup I_{2})}^{} \ln(1 + x_{2i}))} \label{alph2}\end{equation}

Therefore the algorithm can be given as 
\begin{algorithm}
\caption{EM procedure for bivariate Pareto distribution}
\begin{algorithmic}[1]
\STATE Calculate the estimates of $\mu_1$, $\mu_2$, $\sigma_1$ and $\sigma_2$ from marginals of bivariate distribution.
\STATE Fix $I_{0}$, $I_{1}$ and $I_{2}$.
\WHILE{$\Delta Q/Q<tol$}
     \STATE Compute $u^{(i)}_{1}$, $u^{(i)}_{2}$, $w^{(i)}_{1}$, $w^{(i)}_{2}$ from $\alpha^{(i)}_{0}$, $\alpha^{(i)}_{1}$, $\alpha^{(i)}_{2}$.
     \STATE Update $\alpha^{(i+1)}_{0}$, $\alpha^{(i+1)}_{1}$, $\alpha^{(i+1)}_{2}$ using Equation (\ref{alph0}), (\ref{alph1}) and (\ref{alph2}).
     \STATE Calculate $Q$ for the new iterate.
\ENDWHILE
\end{algorithmic}
\label{algo1}
\end{algorithm}

\subsection{Algorithm-\ref{algo2} : Modified Algorithm 1}

 The above algorithm does not work even if all the $I_{0}$, $I_{1}$ and $I_{2}$ are non-empty.  In our previous approach we calculate the estimates of location and scale parameters and construct the above three sets based on the transformed variables using those estimated location and scale parameters.  Therefore it is high likely that there will be almost no element in $I_{0}$ for most of the generated sample.  Consequently it should provide the estimate of $\alpha_{0}$ as zero most of the time.    
 
 We propose to make few modifications in the above EM algorithm.  Since we observe the normalized data with respect to the estimated location and scale parameter, the transformation is not going to provide the distribution of normalized data exactly as Pareto with location zero and scale one. The transformation rather form some distribution close to Pareto with location zero and scale one.  It is very difficult to know the exact distribution.  Approximately even if we assume some Pareto with location parameter near zero and scale parameter near one, calculation of probability that an observation will come from any one of $I_{0}$, $I_{1}$ or $I_{2}$, becomes another problem.  

 To avoid such instances, we assume that information of all $n_{0}$, $n_{1}$ and $n_{2}$ are missing because of small perturbation inherited by estimated location and scale parameters.  In ideal situation we should get EM algorithm steps as described in \ref{algo3}.  Since we have now three more latent information $n_{0}$, $n_{1}$ and $n_{2}$, we try to approximate them and incorporate in the original EM algorithm. 

 It is clear that $(n_{0}, n_{1}, n_{2})$ will jointly follow multinomial distribution with parameters $ n = n_{0} + n_{1} + n_{2}$ and $(\frac{\alpha_{0}}{\alpha_{0} + \alpha_{1} + \alpha_{2}}, \frac{\alpha_{1}}{\alpha_{0} + \alpha_{1} + \alpha_{2}}, \frac{\alpha_{2}}{\alpha_{0} + \alpha_{1} + \alpha_{2}} ). $

  We approximate $\tilde{n_{i}} = n\cdot \frac{\alpha_{i}}{\alpha_{0} + \alpha_{1} + \alpha_{2}} , ~~~~~~ i = 0, 1, 2.$

Therefore M-step involves maximizing (\ref{likelihood}) with respect to $\alpha_{0}$, $\alpha_{1}$, $\alpha_{2}$ at 
\begin{equation} \hat{\alpha}^{(t + 1)}_{0} = \frac{\tilde{n_{0}} + u^{(t)}_{1} \tilde{n_{1}} + w^{(t)}_{1} \tilde{n_{2}}}{\sum_{i \in I_{0}}^{} \ln(1 + x_{i}) + \sum_{i \in I_{2}}^{} \ln(1 + x_{1i}) + \sum_{i \in I_{1}}^{} \ln (1 + x_{2i})}. \label{aalph0}\end{equation}
\begin{equation} \hat{\alpha}^{(t + 1)}_{1} = \frac{(\tilde{n_{1}} + w^{(t)}_{2}\tilde{n_{2}})}{\sum_{i \in I_{0}}^{} \ln(1 + x_{i})  + \sum_{i \in (I_{1} \cup I_{2})}^{} \ln(1 + x_{1i}) } \label{aalph1}\end{equation}
\begin{equation} \hat{\alpha}^{(t + 1)}_{2} = \frac{\tilde{n_{2}} + u^{(t)}_{2} \tilde{n_{1}}}{(\sum_{i \in I_{0}}^{} \ln(1 + x_{i}) +  \sum_{i \in (I_{1} \cup I_{2})}^{} \ln(1 + x_{2i}))} \label{aalph2}\end{equation}

 Finally modified algorithm would be
\begin{algorithm}
\caption{EM procedure for bivariate Pareto distribution}
\begin{algorithmic}[1]
\STATE Calculate the estimates of $\mu_1$, $\mu_2$, $\sigma_1$ and $\sigma_2$ from marginals of bivariate distribution.
\STATE Fix $I_{0}$, $I_{1}$ and $I_{2}$.
\WHILE{$\Delta Q/Q < tol$}
     \STATE Compute $u^{(i)}_{1}$, $u^{(i)}_{2}$, $w^{(i)}_{1}$, $w^{(i)}_{2}$, $\tilde{n_{0}}$, $\tilde{n_{1}}$ and $\tilde{n_{2}}$ from $\alpha^{(i)}_{0}$, $\alpha^{(i)}_{1}$, $\alpha^{(i)}_{2}$.
     \STATE Update $\alpha^{(i+1)}_{0}$, $\alpha^{(i+1)}_{1}$, $\alpha^{(i+1)}_{2}$ using Equation (\ref{aalph0}), (\ref{aalph1}) and (\ref{aalph2}).
     \STATE Calculate $Q$ for the new iterate.
\ENDWHILE
\end{algorithmic}
\label{algo2}
\end{algorithm}

\subsection{Algorithm-\ref{algo3} : Modified Algorithm 2}  In this case we update $\sigma_1$ and $\sigma_2$ along with EM iterations similar to what Asimit et al. has done in their paper.  But the main difference of our update is that it is based on one step ahead gradient descent instead of fixed point iteration on scale parameters.  Surely gradient descent with respect to bivariate likelihood won't work as bivariate likelihood is a discontinuous function with respect to location and scale parameters.  Therefore we estimate scale parameters using density of marginals.  At every iteration we use one step ahead gradient descend of $\sigma_1$ and $\sigma_2$ based on likelihood of its marginal density combined with usual EM steps for other parameters to solve the problem.  The idea is similar to stochastic gradient descend.  We begin the algorithm estimating the sets $I_{0}$, $I_{1}$ and $I_{2}$ which involves location and scale parameters. Given location and scale parameters, we can update previous EM steps of $\alpha_{0}$, $\alpha_{1}$ and $\alpha_{2}$.  EM steps will ensure to take the correct direction of $\alpha_0$, $\alpha_1$ and $\alpha_2$ starting from any values of this three parameters and also it will enable gradient descend steps of $\sigma_1$ and $\sigma_2$ to converge faster starting from any values. 

 This algorithm works even for moderate sample sizes.  However it takes lot of time to converge or roam around the actual value for some really bad sample.  We calculate the estimates both with full iterations until convergence and with at most 2000 iterations. We observe MSEs and the number of iterations in both the cases.  We expect that within 2000 iterations estimated points should be sufficiently closer to the actual value or it should take some value which can be good estimate for starting point of some other optimization algorithms.    

\begin{algorithm}
\caption{EM procedure for bivariate Pareto distribution}
\begin{algorithmic}[2]
\STATE Take the estimates of $\mu_1$, $\mu_2$ by $\mu_1 = \min\{ X_{1i}; i = 1, \cdots, n \}$ and $ \mu_2 = \min\{ X_{2i}; i = 1, \cdots, n \}$.
\STATE Start with some initial choice of $\sigma_1$, $\sigma_2$, $\alpha_{0}$, $\alpha_{1}$, $\alpha_{2}$.
\WHILE{$|\Delta Q/Q| > tol$}
     \STATE Fix $I_{0}$, $I_{1}$ and $I_{2}$ with estimated $\mu_1$, $\mu_2$, $\sigma^{(i)}_1$ and $\sigma^{(i)}_2$. 
     \STATE Compute updates of $\sigma^{(i)}_1$, $\sigma^{(i)}_2$ through one step ahead gradient descend, using $\alpha^{(i)}_{0}$, $\alpha^{(i)}_{1}$, $\alpha^{(i)}_{2}$. 
     \STATE Compute $u^{(i)}_{1}$, $u^{(i)}_{2}$, $w^{(i)}_{1}$, $w^{(i)}_{2}$, $\tilde{n_{0}}$, $\tilde{n_{1}}$ and $\tilde{n_{2}}$ from $\alpha^{(i)}_{0}$, $\alpha^{(i)}_{1}$, $\alpha^{(i)}_{2}$.
     \STATE Update $\alpha^{(i+1)}_{0}$, $\alpha^{(i+1)}_{1}$, $\alpha^{(i+1)}_{2}$ using Equation (\ref{aalph0}), (\ref{aalph1}) and (\ref{aalph2}).
     \STATE Calculate $Q$ for the new iterate.
\ENDWHILE
\end{algorithmic}
\label{algo3}
\end{algorithm}

\subsection{Algorithm-\ref{algo4} : Modified Algorithm 3}  We can make more variations on it.  This is similar to what Asimit et al. has done in their paper.  At each iteration we update $\sigma_1$ and $\sigma_2$ from the marginal density using one step ahead fixed point iteration instead of Gradient decend algorithm and carry on the same EM algorithm.  However approach of Asimit et al is different, therefore EM steps are also differing.  

\begin{algorithm}
\caption{EM procedure for bivariate Pareto distribution}
\begin{algorithmic}[1]
\STATE Take the estimates of $\mu_1$, $\mu_2$ by $\mu_1 = \min\{ X_{1i}; i = 1, \cdots, n \}$ and $ \mu_2 = \min\{ X_{2i}; i = 1, \cdots, n \}$.
\STATE Start with some initial choice of $\sigma_1$, $\sigma_2$, $\alpha_{0}$, $\alpha_{1}$, $\alpha_{2}$.
\WHILE{$|\Delta Q/Q| > tol$}
     \STATE Fix $I_{0}$, $I_{1}$ and $I_{2}$ with estimated $\mu_1$, $\mu_2$, $\sigma^{(i)}_1$ and $\sigma^{(i)}_2$. 
     \STATE Compute one step ahead update of $\sigma^{(i)}_1$, $\sigma^{(i)}_2$ through fixed point iteration, using $\alpha^{(i)}_{0}$, $\alpha^{(i)}_{1}$, $\alpha^{(i)}_{2}$.  
     \STATE Compute $u^{(i)}_{1}$, $u^{(i)}_{2}$, $w^{(i)}_{1}$, $w^{(i)}_{2}$, $\tilde{n_{0}}$, $\tilde{n_{1}}$ and $\tilde{n_{2}}$ from $\alpha^{(i)}_{0}$, $\alpha^{(i)}_{1}$, $\alpha^{(i)}_{2}$.
     \STATE Update $\alpha^{(i+1)}_{0}$, $\alpha^{(i+1)}_{1}$, $\alpha^{(i+1)}_{2}$ using Equation (\ref{aalph0}), (\ref{aalph1}) and (\ref{aalph2}).
     \STATE Calculate $Q$ for the new iterate.
\ENDWHILE
\end{algorithmic}
\label{algo4}
\end{algorithm}
      
\subsection{Algorithm-\ref{algo5} : Modified Algorithm 4}  This variation is of two fold.  We update $\sigma_1$ and $\sigma_2$ from the marginal density until convergence using fixed point iteration instead of Gradient decent algorithm and carry forward the same EM algorithm.  However at the time of using EM iterations, we update the $\sigma_1$ and $\sigma_2$ using one step ahead fixed point iteration along with other parameters.  The detailed algorithmic steps are shown below.   
      
\begin{algorithm}
\caption{EM procedure for bivariate Pareto distribution}
\begin{algorithmic}[1]
\STATE Start with some initial choice of $\sigma_1$, $\sigma_2$, $\alpha_{0}$, $\alpha_{1}$, $\alpha_{2}$.
\WHILE{$|(\Delta\sigma)| > tol$}
\STATE Compute one step ahead fixed point iteration of $\sigma^{(i + 1)}_1$, $\sigma^{(i + 1)}_2$, using $\sigma^{(i)}_1$, $\sigma^{(i)}_2$, $\alpha^{(i)}_{0}$, $\alpha^{(i)}_{1}$, $\alpha^{(i)}_{2}$.
\STATE Compute $u^{(i)}_{1}$, $u^{(i)}_{2}$, $w^{(i)}_{1}$, $w^{(i)}_{2}$, $\tilde{n_{0}}$, $\tilde{n_{1}}$ and $\tilde{n_{2}}$ from $\alpha^{(i)}_{0}$, $\alpha^{(i)}_{1}$, $\alpha^{(i)}_{2}$.
\STATE Update $\alpha^{(i+1)}_{0}$, $\alpha^{(i+1)}_{1}$, $\alpha^{(i+1)}_{2}$ using Equation (\ref{aalph0}), (\ref{aalph1}) and (\ref{aalph2}). 
\ENDWHILE
\WHILE{$|\Delta Q/Q| > tol$}
     \STATE Fix $I_{0}$, $I_{1}$ and $I_{2}$ with last updated estimates of $\mu_1$, $\mu_2$, $\sigma^{(i)}_1$, $\sigma^{(i)}_2$, $\alpha^{(i)}_{0}$, $\alpha^{(i)}_{1}$ and $\alpha^{(i)}_{2}$.
     \STATE Compute updates of $\sigma^{(i + 1)}_1$, $\sigma^{(i + 1)}_2$, using $\alpha^{(i)}_{0}$, $\alpha^{(i)}_{1}$, $\alpha^{(i)}_{2}$. 
     \STATE Compute $u^{(i)}_{1}$, $u^{(i)}_{2}$, $w^{(i)}_{1}$, $w^{(i)}_{2}$, $\tilde{n_{0}}$, $\tilde{n_{1}}$ and $\tilde{n_{2}}$ from $\alpha^{(i)}_{0}$, $\alpha^{(i)}_{1}$, $\alpha^{(i)}_{2}$.
     \STATE Update $\alpha^{(i+1)}_{0}$, $\alpha^{(i+1)}_{1}$, $\alpha^{(i+1)}_{2}$ using Equation (\ref{aalph0}), (\ref{aalph1}) and (\ref{aalph2}).
     \STATE Calculate $Q$ for the new iterate. 
\ENDWHILE
\end{algorithmic}
\label{algo5}
\end{algorithm}

\section{Numerical Results}  

  We use package R 3.2.3 to perform the estimation procedure.  All the programs will be available from author on request.  First we take four different sets of parameters and observe the average iteration calculated over different sample sizes. We take our sample size as $n = 150, 250, 350, 450$.   We can find the results in Table-\ref{table-AI} based on 1000 replications.  Tolerance limit (denoted as "tol") of stopping criteria is taken as 0.00001. We have used stopping criteria as absolute value of likelihood changes with respect to previous Qseudo-likelihood at each iteration.  Results shown in Table-\ref{table-AI} provide average iteration by six different approaches.  They are (a) Modified Approach 1 (b) Modified Approach 2 which we truncate after 2000 iterations and denote this as Modified Approach 2 (T)  (c) Modified Approach 2 without truncating the iteration and denote this as Modified Approach 2 (WT) (d) Modified Approach 3 (e) Modified Approach 4 (f) Asimit's approach. We choose four different sets of parameters to compare the results. 
These are $\mu_1 = 0, \mu_2 = 0, \sigma_1 = 1, \sigma_2 = 0.5, \alpha_0 = 1, \alpha_1 = 0.3, \alpha_2 = 1.4$; ~~~
 $\mu_1 = 1, \mu_2 = 2, \sigma_1 = 0.4, \sigma_2 = 0.5, \alpha_0 = 2, \alpha_1 = 1.2, \alpha_2 = 1.4$;~~~
 $\mu_1 = 0, \mu_2 = 0, \sigma_1 = 1.4, \sigma_2 = 0.5, \alpha_0 = 1, \alpha_1 = 1, \alpha_2 = 1.4$;~~~
 $\mu_1 = 0, \mu_2 = 0, \sigma_1 = 1.4, \sigma_2 = 0.5, \alpha_0 = 2, \alpha_1 = 0.4, \alpha_2 = 0.5$.~~~
  The proposed EM algorithms work for any initial value.  However estimates of $\mu_1$ and $\mu_2$ are always $\hat{\mu}_1 = \min\{ X_{1i}; 1, \cdots, n\}$ and $\hat{\mu}_2 = \min\{ X_{2i}; 1, \cdots, n\}$ respectively.  Original paper of Asimit et al. uses different stopping criteria and results are provided for large sample sizes.  To compare our results, we keep stopping criteria and sample size same across all algorithms.  

 Average estimates are provided in Table-\ref{Table-AE} when samples are simulated from bivariate pareto with parameters $\mu_1 = 0, \mu_2 = 0, \sigma_1 = 1, \sigma_2 = 0.5, \alpha_0 = 1, \alpha_1 = 0.3, \alpha_2 = 1.4$.  Results of average estimates are shown only by two best approaches (best in the sense of mininum average iteration) i.e. Modified Approach 1 and Modified Approach 4. Table-\ref{table-MSE1} and Table-\ref{table-MSE2} show the MSEs for all procedures where samples are generated from the following parameter sets : $\mu_1 = 0, \mu_2 = 0, \sigma_1 = 1, \sigma_2 = 0.5, \alpha_0 = 1, \alpha_1 = 0.3, \alpha_2 = 1.4$.  MSE is an important criteria for selecting the best algorithm. Therefore MSEs are calculated for all the procedures.  We also calculate parametric bootstrap confidence interval.  We simulate 1000 samples and estimate of the parameters based on the simulated samples.  We use them to get the 95\% confidence interval by calculating 0.025 and 0.975 sample quantile points of estimated parameters.  Table-\ref{tab:para-boot1} to Table-\ref{tab:para-boot5} carry the information related to parametric bootstrap confidence intervals.   

\noindent{\underline{Important Comments :}}  

\begin{enumerate}

\item Mean square error for $\alpha_2$ is little higher for all methods.

\item MSEs are more or less same for all methods. In the above result we see Modified Approach 1 provides minimum MSE among all procedures.  For large sample size Modified Approach 1 comes out as winner for any chosen parameter sets. Performance of Modified Approach 2 with truncation is also worth mentioning, as it provides second best performance with respect to minimum MSE.

\item Modified Approach 4 is the best among other approaches in terms of average iteration. Modified Approach 1 appears to be the second best performer. However average iteration for Modified Approach 2 with truncation is much higher as compared to Modified Approach 1 and 4.

\item As expected, average iteration is very high in case of Modified Approach 2 (WT).  

\item If we compare the algorithms based on minimum MSE and average iteration together, Modified approach 1 stands either winner or closer to the winner. 

\end{enumerate}

\begin{table}[!h]
\begin{scriptsize}
\begin{center}
\begin{tabular}{|c|c|c|c|c|c|c|}\hline
parameter set  &  & n  &  150 & 250 & 350 & 450  \\
 &    &    &     &     &  &   \\ \hline
$\mu_1 = 0$, $\mu_2 = 0$ & The average & Modified  &  244 & 231 & 223 & 220 \\ 
$\sigma_1 = 1$ $\sigma_2 = 0.5$ & number & Approach 1 &     &     &   &   \\
$\alpha_0 = 1$, $\alpha_1 = 0.3$,  & of Iterations & Modified & 3402 & 2163 & 1542 & 1220 \\
$\alpha_2 = 1.4$ & (AI)   & Approach 2 (WT) &     &     &   &   \\ 
 &  & Modified & 1771 & 1631 & 1414 & 1194  \\ 
 &  & Approach 2 (T) &  &  &  & \\ 
 &  & Modified &  266 & 249 & 238 & 234\\
 &  & Approach 3 & & & &\\ 
 &  & Modified & 191 & 162 & 151 & 147\\
 &  & Approach 4 &  &  &   &   \\ 
 &  & Asimit's Approach & 1002  & 945 & 909 & 902 \\ 
 &  &    &     &     &  &   \\ \hline
$\mu_1 = 1$, $\mu_2 = 2$ & The average  & Modified & 174 & 165 & 159 & 157 \\
$\sigma_1 = 0.4$ $\sigma_2 = 0.5$ & number & Approach 1  &    &     &     &   \\
$\alpha_0 = 2$, $\alpha_1 = 1.2$, & of Iterations & Modified & 7734 & 4137 & 2428 & 1633 \\
$\alpha_2 = 1.4$ & (AI) & Approach 2 (WT) &  &     &   & \\ 
 &  & Modified &  1704  &  1553  &  1378  &  1184  \\ 
 &  & Approach 2 (T) &    &     &   &   \\ 
 &  & Modified & 764 & 551 & 458 & 424 \\
 &  & Approach 3 &  &  &   &   \\ 
 &  & Modified &  243 & 185 & 153 & 128\\
 &  & Approach 4 &  &  &   &   \\ 
 &  & Asimit's Approach & 2345 & 1747 & 1464 & 1309 \\ 
 &  &    &     &     &  &   \\ \hline
 $\mu_1 = 0$, $\mu_2 = 0$  & The average  & Modified  & 168  & 155 & 150 & 147 \\
 $\sigma_1 = 1.4$ $\sigma_2 = 0.5$ & number & Approach 1  &     &     &   &   \\
$\alpha_0 = 1$, $\alpha_1 = 1$, & of Iterations & Modified  & 7230 & 4372 & 3077 & 2423 \\
$\alpha_2 = 1.4$ & (AI)  & Approach 2 (WT)  &     &     &   &   \\ 
 &  & Modified  &  1617 &  1738 & 1732 & 1697 \\ 
 &  & Approach 2 (T) &    &     &   &   \\ 
 &  & Modified & 265 & 237 & 220 & 209 \\
 &  & Approach 3 &  &  &   &   \\ 
 &  & Modified & 123 & 101 & 91 & 83\\
 &  & Approach 4 &  &  &   &   \\ 
 &  & Asimit's Approach & 897 & 547 & 484 & 453 \\ 
 &  &    &     &     &  &   \\ \hline
 $\mu_1 = 0$, $\mu_2 = 0$ & The average  & Modified  & 138 & 129 & 124 & 122 \\
 $\sigma_1 = 1.4$, $\sigma_2 = 0.5$ & number & Approach 1  &     &     &   &   \\
 $\alpha_0 = 2$, $\alpha_1 = 0.4$, & of Iterations & Modified  &  2172 & 2707 & 2237 & 1821 \\
$\alpha_2 = 0.5$ & (AI) & Approach 2 (WT)  &     &     &   &  \\
 &  & Modified   &  397  &  591  & 773  &  881 \\ 
&  & Approach 2 (T)  &     &     &   &   \\ 
&  & Modified & 330 & 261 & 238 & 219 \\
 &  & Approach 3 &  &  &   &   \\ 
&  & Modified & 117 & 100 & 96 & 95\\
 &  & Approach 4 &  &  &   &   \\ 
&  & Asimit's Approach &  1145  & 977 & 919 & 875 \\ 
&  &    &     &     &  &   \\ \hline 
\end{tabular}  
\caption{Average number of iterations (AI) by six approaches: (a) Modified Approach 1 (b) Modified Approach 2 (WT) (c) Modified Approach 2 (T) (d) Modified Approach 3 (e) Modified Approach 4 (f) Asimit's Approach}  
\label{table-AI}
\end{center}
\end{scriptsize}
\end{table}

\begin{table}[!h]
\begin{scriptsize}
\begin{center}
\begin{tabular}{|c|c|c|c|c|c|c|c|c|}\hline
 & parameters & $\mu_1$ & $\mu_2$ & $\sigma_1$ & $\sigma_2$ & $\alpha_0$ & $\alpha_1$ & $\alpha_2$ \\
n & &  &    &     &     &  &  & \\ \hline 
  
150  & Average (Mod. 1) &   &    &     &     &  &  & \\
     & estimates & 0.0056 & 0.0014 & 0.9031 & 0.5154 & 0.8993 & 0.3379 & 1.5660\\
     & (AE) & &   &     &     &   &  & \\       
     & Average (Mod. 4)&  &    &     &     &  &  & \\
     & estimates & 0.0056 & 0.0014 & 1.0521 & 0.5755 & 0.9372 & 0.4122 & 1.7416\\
     & (AE) & &   &     &     &   &  & \\        

250 & Average (Mod. 1)& &  &    &     &     &  &   \\
    & Estimates & 0.0031 & 0.0008 & 0.8943 & 0.4989 & 0.9013 & 0.3226 & 1.5046\\  
    & (AE) & &   &     &     &   &  & \\    
    & Average (Mod. 4) &  &    &     &     &  &  & \\
    & estimates & 0.0031 & 0.0008 & 1.0386 & 0.5457 & 0.9408 & 0.3909 & 1.6292\\
    & (AE) & &   &     &     &   &  & \\       
 
350 & Average (Mod. 1) & &  &    &     &     &  &   \\
    & Estimates & 0.0022 & 0.0006 & 0.8854 & 0.4878 & 0.9086 & 0.3107 & 1.4550 \\
    & (AE) & &   &     &     &   &  & \\  
    & Average (Mod. 4) &  &    &     &     &  &  & \\
    & estimates & 0.0022 & 0.0006 & 1.0252 & 0.5283 & 0.9497 & 0.3745 & 1.5543\\
    & (AE) & &   &     &     &   &  & \\       

450 & Average (Mod. 1) &  &  &    &     &     &  &   \\
    & Estimates & 0.0017 & 0.0005 & 0.8719 & 0.4799 & 0.9099 & 0.2955 & 1.4270 \\
    & (AE) & &   &     &     &   &  & \\     
    & Average (Mod. 4) &  &    &     &     &  &  & \\
    & estimates & 0.0017 & 0.0005 & 1.0077 & 0.5187 & 0.9492 & 0.3586 & 1.5221\\
    & (AE) & &   &     &     &   &  & \\ \hline      
\end{tabular}  
\caption{The average estimates (AE) for $\mu_1 = 0$, $\mu_2 = 0$, $\sigma_1 = 1$, $\sigma_2 = 0.5$, $\alpha_0 = 1$, $\alpha_1 = 0.3$ and $\alpha_2 = 1.4$ through two best approches i.e. Modified Approach 1 and Modified Approach 4 \label{Table-AE}}  
\end{center}
\end{scriptsize}
\end{table}

\begin{table}[!h]
\begin{scriptsize}
\begin{center}
\begin{tabular}{|c|c|c|c|c|}\hline
 & parameters & $\mu_1$ & $\mu_2$ & $\sigma_1$\\  
 &     &  &  & \\ \hline   
n=150 & MSE (Mod. 1) & 6.14e-05 & 3.69e-06 & 0.0832 \\  
     & MSE (Mod. 2 (T)) & 0.0000614 & 0.0000037 & 0.0543 \\  
     & MSE (Mod. 2 (WT)) & 0.0000614 & 0.0000370 & 0.1043 \\ 
     & MSE (Mod. 3) & 0.0000614 & 0.0000037 & 0.1216 \\  
     & MSE (Mod. 4) & 0.0000614 & 0.0000037 & 0.1179 \\  
     & MSE (Asimit) & 0.0000614 & 0.0000037 & 0.1218 \\  \hline       
 n = 150 & $\sigma_2$ & $\alpha_0$ & $\alpha_1$ & $\alpha_2$ \\
MSE (Mod. 1) & 0.0477 & 0.0402 & 0.0375  & 0.5178 \\  
MSE (Mod. 2 (T)) & 0.0368 & 0.0386 & 0.0342 & 0.4154 \\
MSE (Mod. 2 (WT)) & 0.0682 & 0.0433 & 0.0609 & 0.7969 \\  
MSE (Mod. 3) &  0.1028 & 0.0439 & 0.0737 & 1.2035 \\ 
MSE (Mod. 4) & 0.1090 & 0.0429 & 0.0691 & 1.2536 \\ 
MSE (Asimit) & 0.1011 & 0.0433 & 0.0726 & 1.1860 \\ \hline       
  & parameters & $\mu_1$ & $\mu_2$ & $\sigma_1$\\  
   &     &  &  & \\ \hline   
n = 250 & MSE (Mod. 1) & 0.000020  & 1.4757e-06 & 0.0581 \\   
    & MSE (Mod. 2 (T)) & 0.0000614 & 0.00000370 & 0.0543 \\  
    & MSE (Mod. 2 (WT)) & 0.0000198 & 0.0000014 & 0.0682 \\  
    & MSE (Mod. 3) & 0.00001978 & 0.00000147 & 0.0745 \\  
    & MSE (Mod. 4) & 0.00001978 & 0.00000147 & 0.0724 \\  
    & MSE (Asimit) & 0.00001978 & 0.00000148 & 0.0746 \\  
n = 250  & $\sigma_2$ & $\alpha_0$ & $\alpha_1$ & $\alpha_2$ \\
MSE (Mod. 1) & 0.0278  & 0.0296  & 0.0205 & 0.2817 \\   
MSE (Mod. 2 (T)) & 0.0368 & 0.0386 & 0.0342 & 0.4153 \\ 
MSE (Mod. 2 (WT)) & 0.0410 & 0.0294 & 0.0386 & 0.4441 \\ 
MSE (Mod. 3) & 0.0461 & 0.0296 & 0.0431 & 0.5033 \\
MSE (Mod. 4) & 0.0460 & 0.0289 & 0.0401 & 0.4978 \\ 
MSE (Asimit) & 0.0462 & 0.0292 & 0.0427 & 0.5638 \\ \hline        
\end{tabular}  
\caption{Mean Square Error (MSE) through all approaches when samples are generated from $\mu_1 = 0$, $\mu_2 = 0$, $\sigma_1 = 1$, $\sigma_2 = 0.5$, $\alpha_0 = 1$, $\alpha_1 = 0.3$ and $\alpha_2 = 1.4$ \label{table-MSE1}}  
\end{center}
\end{scriptsize}
\end{table}

\begin{table}[!h]
\begin{scriptsize}
\begin{center}
\begin{tabular}{|c|c|c|c|c|}\hline
 & parameters & $\mu_1$ & $\mu_2$ & $\sigma_1$ \\ 
 &  &  &  & \\ \hline   
n = 350 & MSE (Mod. 1) & 9.236e-06 & 6.992e-07 & 0.0427 \\ 
    & MSE (Mod. 2 (T)) & 0.00000197 & 0.000000699 & 0.0389\\ 
    & MSE (Mod. 2 (WT)) & 0.00000924 & 0.000000699 & 0.0428 \\ 
    & MSE (Mod. 3) & 0.00000923 & 0.000000699 & 0.0456 \\ 
    & MSE (Mod. 4) & 0.00000924 & 0.000000699 & 0.0440 \\ 
    & MSE (Asimit) & 0.00000923 & 0.000000699 & 0.0454 \\  
n=350 & $\sigma_2$ & $\alpha_0$ & $\alpha_1$ & $\alpha_2$ \\
MSE (Mod. 1) & 0.0155     & 0.0211      & 0.0143    & 0.1475  \\   
MSE (Mod. 2 (T)) &  0.02001   & 0.01907    & 0.02573    & 0.2068 \\
MSE (Mod. 2 (WT)) &  0.0213    & 0.0192     & 0.0280     & 0.2204 \\  
MSE (Mod. 3) &  0.0221    & 0.0194     & 0.0302     & 0.2290 \\ 
MSE (Mod. 4) &  0.0220    & 0.0188     & 0.0275     & 0.2252 \\  
MSE (Asimit) &   0.0222   & 0.0194     & 0.0301     & 0.2290 \\\hline  

 & parameters & $\mu_1$ & $\mu_2$ & $\sigma_1$ \\ 
 &  &  &  & \\ \hline          
n = 450 & MSE (Mod. 1) & 5.589e-06 & 4.398e-07 & 0.0366 \\  
    & MSE (Mod. 2 (T)) & 0.0000055 & 0.000000440 & 0.0283\\ 
    & MSE (Mod. 2 (WT)) & 0.00000559 & 0.00000044 & 0.0288 \\ 
    & MSE (Mod. 3) & 0.00000559 & 0.00000047 & 0.03008 \\ 
    & MSE (Mod. 4) & 0.00000558 & 0.000000439 & 0.0292 \\ 
    & MSE (Asimit) & 0.00000559 & 0.000000440 & 0.0300 \\ 
n=450 & $\sigma_2$ & $\alpha_0$ & $\alpha_1$ & $\alpha_2$ \\
MSE (Mod. 1) & 0.0107 & 0.0176  & 0.0097 & 0.1140   \\  
MSE (Mod. 2 (T)) & 0.0143 & 0.015 & 0.0181 & 0.1655 \\  
MSE (Mod. 2 (WT)) & 0.0147 & 0.0151 & 0.0184 & 0.1717 \\
MSE (Mod. 3) & 0.0151 & 0.0152 & 0.0196 & 0.1759 \\ 
MSE (Mod. 4) & 0.0150 & 0.0148 & 0.0179 & 0.1736 \\  
MSE (Asimit) & 0.0151 & 0.0151 & 0.0195 & 0.1756 \\ \hline      
\end{tabular}  
\caption{Mean Square Error (MSE) through all approaches when samples are generated from $\mu_1 = 0$, $\mu_2 = 0$, $\sigma_1 = 1$, $\sigma_2 = 0.5$, $\alpha_0 = 1$, $\alpha_1 = 0.3$ and $\alpha_2 = 1.4$ \label{table-MSE2}}  
\end{center}
\end{scriptsize}
\end{table}

\begin{table}[!h]
\begin{small}
\begin{center}
\begin{tabular}{|c|c|c|c|c|}\hline
 n &  150 & 250 & 350 & 450  \\
 parameters&  &    &     &   \\ \hline 
 $\mu_1$ &  [0.0001, 0.0193] & [0.000077, 0.0125] & [0.00007, 0.0079] & [0.00004, 0.0062] \\
$\mu_2$ &  [0.000034, 0.0050] & [0.000027, 0.0034] & [0.00002, 0.0022] & [0.00001, 0.0017] \\
$\sigma_1$ &  [0.5112, 1.4976] & [0.5513, 1.3754] & [0.6016, 1.2959] & [0.6257, 1.1722] \\ 
$\sigma_2$ &  [0.2361, 1.0501] & [0.2796, 0.9011] & [0.3067, 0.7946] & [0.3183, 0.7207] \\
$\alpha_0$ & [0.6103, 1.2854] & [0.6682, 1.1940] & [0.7004, 1.1586] & [0.7373, 1.1197]  \\
$\alpha_1$ &  [0.0974, 0.8052] & [0.1172, 0.6801] & [0.1206, 0.5963] & [0.1241, 0.5377] \\
$\alpha_2$ &  [0.7692, 3.3941] & [0.8701, 2.8723] & [0.9447, 2.3398] & [0.9772, 2.2408] \\ \hline
\end{tabular}  
\caption{Parametric Bootstrap confidence interval for $\mu_1 = 0$, $\mu_2 = 0$, $\sigma_1 = 1$, $\sigma_2 = 0.5$, $\alpha_0 = 1$, $\alpha_1 = 0.3$ and $\alpha_2 = 1.4$ in Modified Approach 1 }
\label{tab:para-boot1}  
\end{center}
\end{small}
\end{table}

\begin{table}[!h]
\begin{small}
\begin{center}
\begin{tabular}{|c|c|c|c|c|}\hline
 n &  150 & 250 & 350 & 450  \\
 parameters&  &    &     &   \\ \hline 
 $\mu_1$ &  [0.0001, 0.01929] & [0.000077, 0.01248] & [0.000067, 0.00792] & [0.0000449, 0.00619] \\
$\mu_2$ &   [0.000034, 0.004952] & [0.0000265, 0.00337] & [0.0000203, 0.002219] & [0.0000155, 0.00175] \\
$\sigma_1$ &  [0.5872, 1.4421] & [0.6316, 1.4810] & [0.6921, 1.4760] & [0.71373, 1.3677] \\ 
$\sigma_2$ &  [0.2497, 0.9518] & [0.2876, 0.9674] & [0.3277, 0.8631] & [0.33937, 0.79859] \\
$\alpha_0$ &  [0.6146, 1.3454] & [0.6684, 1.2655] & [0.7089, 1.2288] & [0.75384, 1.17617] \\
$\alpha_1$ &  [0.1340, 0.8017] & [0.1608, 0.7578] & [0.1772, 0.7227] & [0.19451, 0.66522] \\
$\alpha_2$ &  [0.7949, 3.0734] & [0.9092, 3.0089] & [0.9932, 2.5686] & [1.03731, 2.455004] \\ \hline
\end{tabular}  
\caption{Parametric Bootstrap confidence interval for $\mu_1 = 0$, $\mu_2 = 0$, $\sigma_1 = 1$, $\sigma_2 = 0.5$, $\alpha_0 = 1$, $\alpha_1 = 0.3$ and $\alpha_2 = 1.4$ in Modified Approach 2 (T)}  
\label{tab:para-boot2}
\end{center}
\end{small}
\end{table}

\begin{table}[!h]
\begin{small}
\begin{center}
\begin{tabular}{|c|c|c|c|c|}\hline
 n &  150 & 250 & 350 & 450  \\
 parameters&  &    &     &    \\ \hline 
 $\mu_1$ & [0.000108, 0.01929] & [0.000077, 0.01248] & [0.000067, 0.00791] & [0.000044, 0.0061] \\
$\mu_2$ & [0.0000343, 0.00495] & [0.0000265, 0.00337] & [0.0000203, 0.0022] & [0.000015, 0.0017] \\
$\sigma_1$ & [0.5872, 1.7322] & [0.6316, 0.1.6122] & [0.69207, 1.5152] & [0.7137, 1.3751]\\ 
$\sigma_2$ & [0.2496, 1.1774] & [0.2877, 1.0454] & [0.3277, 0.8711] & [0.3394, 0.7986] \\
$\alpha_0$ & [0.5967, 1.3852] & [0.6684, 1.2719] & [0.7089, 1.2281] & [0.7538, 1.1762] \\
$\alpha_1$ & [0.1346, 0.9461] & [0.1585, 0.8536] & [0.1773, 0.7503] & [0.1945, 0.6796] \\
$\alpha_2$ & [0.7951, 3.7727] & [0.9093, 3.2908] & [0.9932, 2.5707] & [1.0373, 2.4596] \\ \hline
\end{tabular}  
\caption{Parametric Bootstrap confidence interval for $\mu_1 = 0$, $\mu_2 = 0$, $\sigma_1 = 1$, $\sigma_2 = 0.5$, $\alpha_0 = 1$, $\alpha_1 = 0.3$ and $\alpha_2 = 1.4$ in Modified Approach 2 (WT)}  
\label{tab:para-boot3}
\end{center}
\end{small}
\end{table}

\begin{table}[!h]
\begin{small}
\begin{center}
\begin{tabular}{|c|c|c|c|c|}\hline
 n  & 150 & 250 & 350 & 450  \\
 parameters&  &    &     &   \\ \hline 
 $\mu_1$ & [0.000108, 0.019] & [0.000077, 0.0125] & [0.000067, 0.0079] & [0.000045, 0.0062] \\
$\mu_2$ & [0.000034, 0.0049] & [0.000026, 0.00337] & [0.0000203, 0.0022] & [0.000016, 0.0017]  \\
$\sigma_1$ & [0.5846, 1.1874] & [0.6355, 1.6567] & [0.6956, 1.5428] & [0.7161, 1.3938]  \\ 
$\sigma_2$ & [0.2406, 1.2967] & [0.2875, 1.0817] & [0.3277, 0.8795] & [0.3393, 0.8067]  \\
$\alpha_0$ & [0.5921, 1.4047] & [0.6656, 1.2716] & [0.7075, 1.2319] & [0.7520, 1.1790]  \\
$\alpha_1$ & [0.1378, 0.9898] & [0.1621, 0.8843] & [0.1798, 0.7653] & [0.1946, 0.6894]  \\
$\alpha_2$ & [0.8007, 4.127] & [0.9103, 3.4069] & [0.9941, 2.6199] & [1.0376, 2.4805]  \\ \hline
\end{tabular}  
\caption{Parametric Bootstrap confidence interval for $\mu_1 = 0$, $\mu_2 = 0$, $\sigma_1 = 1$, $\sigma_2 = 0.5$, $\alpha_0 = 1$, $\alpha_1 = 0.3$ and $\alpha_2 = 1.4$ in Modified Approach 3}  
\label{tab:para-boot4}
\end{center}
\end{small}
\end{table}

\begin{table}[!h]
\begin{small}
\begin{center}
\begin{tabular}{|c|c|c|c|c|}\hline
 n  & 150 & 250 & 350 & 450  \\
 parameters&  &    &     &  \\ \hline 
 $\mu_1$ & [0.0001, 0.0193] & [0.000077, 0.0125] & [0.000067, 0.00791] & [0.000044, 0.0062] \\
$\mu_2$ &  [0.000034, 0.0050] & [0.000026, 0.00337] & [0.0000203, 0.0022] & [0.000015, 00017] \\
$\sigma_1$ & [0.5821, 1.7914] & [0.6332, 1.6449] & [0.6936, 1.5209] & [0.7136, 1.3763] \\
$\sigma_2$ & [0.2469, 1.2961] & [0.2874, 1.0815] & [0.3246, 0.8798] & [0.3373, 0.8037] \\
$\alpha_0$ & [0.6102, 1.4071] & [0.6792, 1.2729] & [0.7124, 1.2345] & [0.7562, 1.1791] \\
$\alpha_1$ & [0.1261, 0.9899] & [0.1518, 0.8474] & [0.1737, 0.7367] & [0.1892, 0.6740] \\
$\alpha_2$ & [0.7934, 4.1231] & [0.9044, 3.4036] & [0.9831, 2.6059] & [1.0284, 2.4709] \\ \hline
\end{tabular}  
\caption{Parametric Bootstrap confidence interval for $\mu_1 = 0$, $\mu_2 = 0$, $\sigma_1 = 1$, $\sigma_2 = 0.5$, $\alpha_0 = 1$, $\alpha_1 = 0.3$ and $\alpha_2 = 1.4$ in Modified Approach 4}  
\label{tab:para-boot5}
\end{center}
\end{small}
\end{table}

\section{Data Analysis}

  We analyze a data set on the indemnity payments (Loss) and allocated loss adjustment expense (ALAE) relating to 1500
general liability claims from insurance companies are available in the R package evd (\cite{Stephenson:2015}). From \cite{FalkGuillou:2008}, we know that peak over threshold method on random variable U provides polynomial generalized Pareto distribution for any $x_{0}$ with $1 + log(G(x_{0})) \in (0, 1)$ i.e.  $P(U > t x_{0} | U > x_{0}) = t^{-\alpha}, ~~t \geq 1$ where $G(\cdot)$ is the distribution function of $U$.  We choose appropriate $t$ and $x_{0}$ so that data should behave like near Pareto distribution. 
 
\begin{figure}[!ht]
 \begin{center}
  \subfigure[$\xi_{1}$]{\epsfig{file = 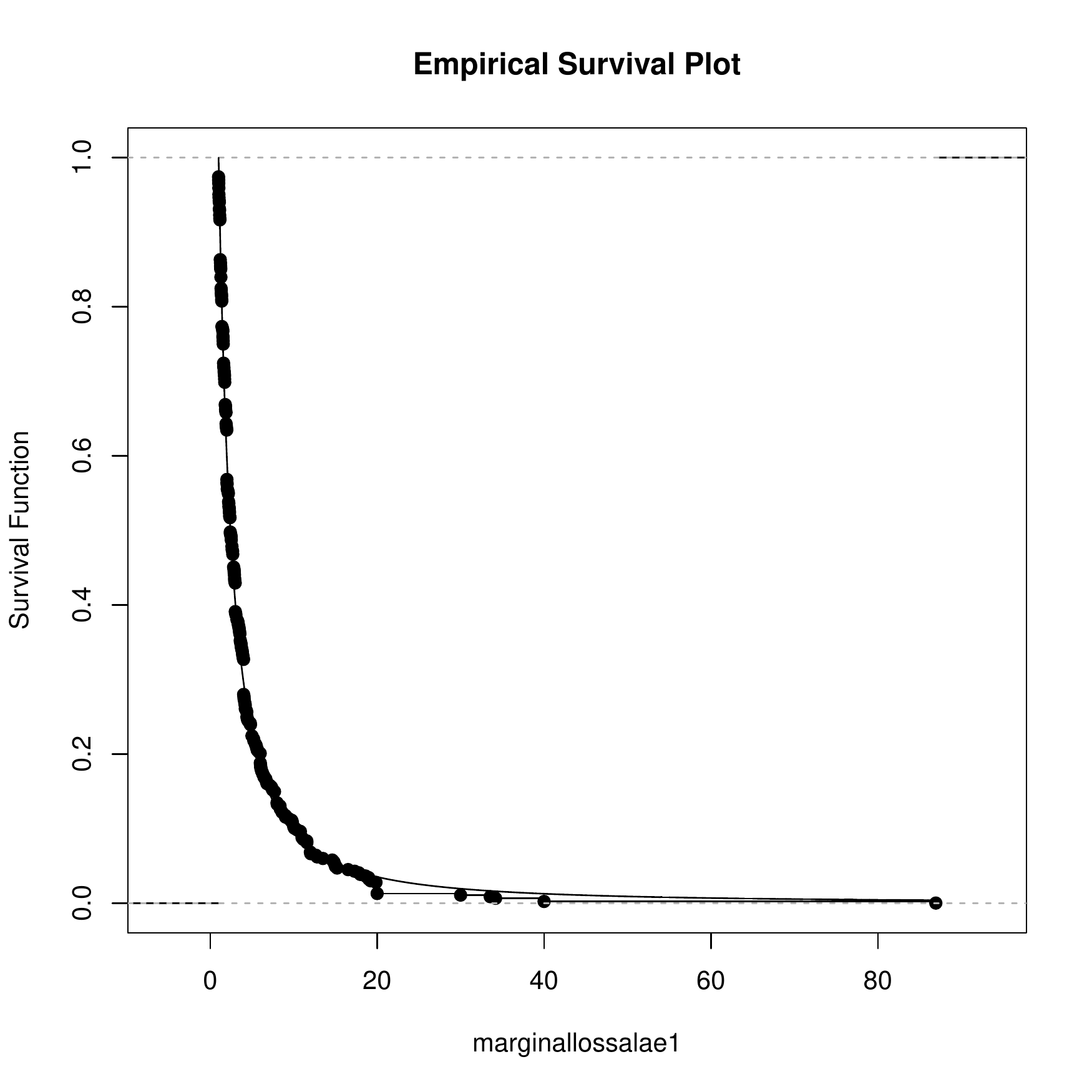, width = 4cm}}
  \subfigure[$\xi_{2}$]{\epsfig{file = 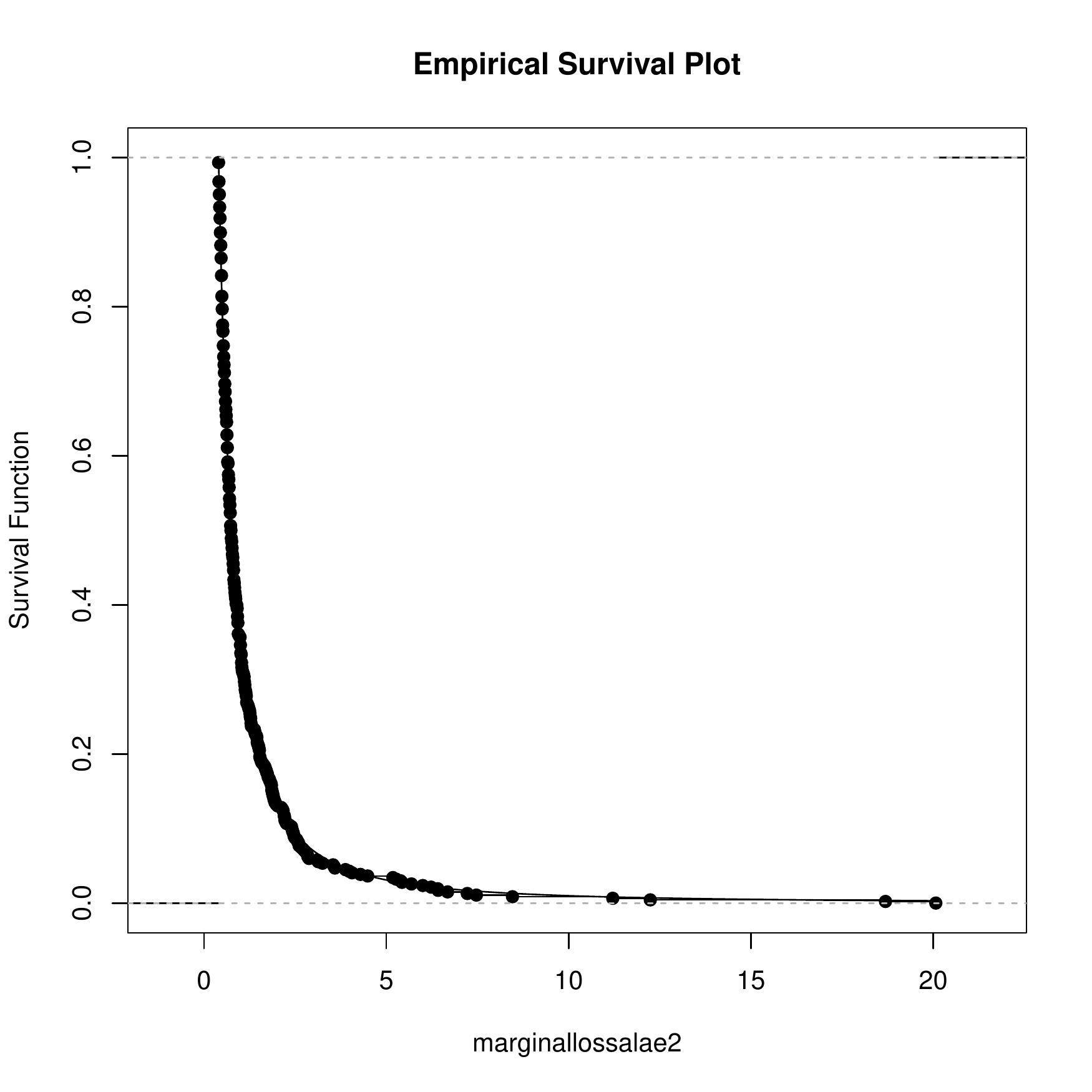, width = 4cm}}\\
\caption{Survival plots for two marginals of the transformed dataset \label{fig3}}
\end{center}
\end{figure}

 We assume the data to follow near equal to singular Marshall Olkin bivariate Pareto and try to verify our assumption. We fit the empirical survival functions with the marginals of this bivariate Pareto whose parameters can be obtained from the EM algorithm that we have developed.  Figure-\ref{fig3} shows a good fit for both the marginals.  The data analysis is performed based on sample size 1500.  

\begin{figure}[ht!]
 \begin{center}
  \includegraphics[width = 0.45\textwidth]{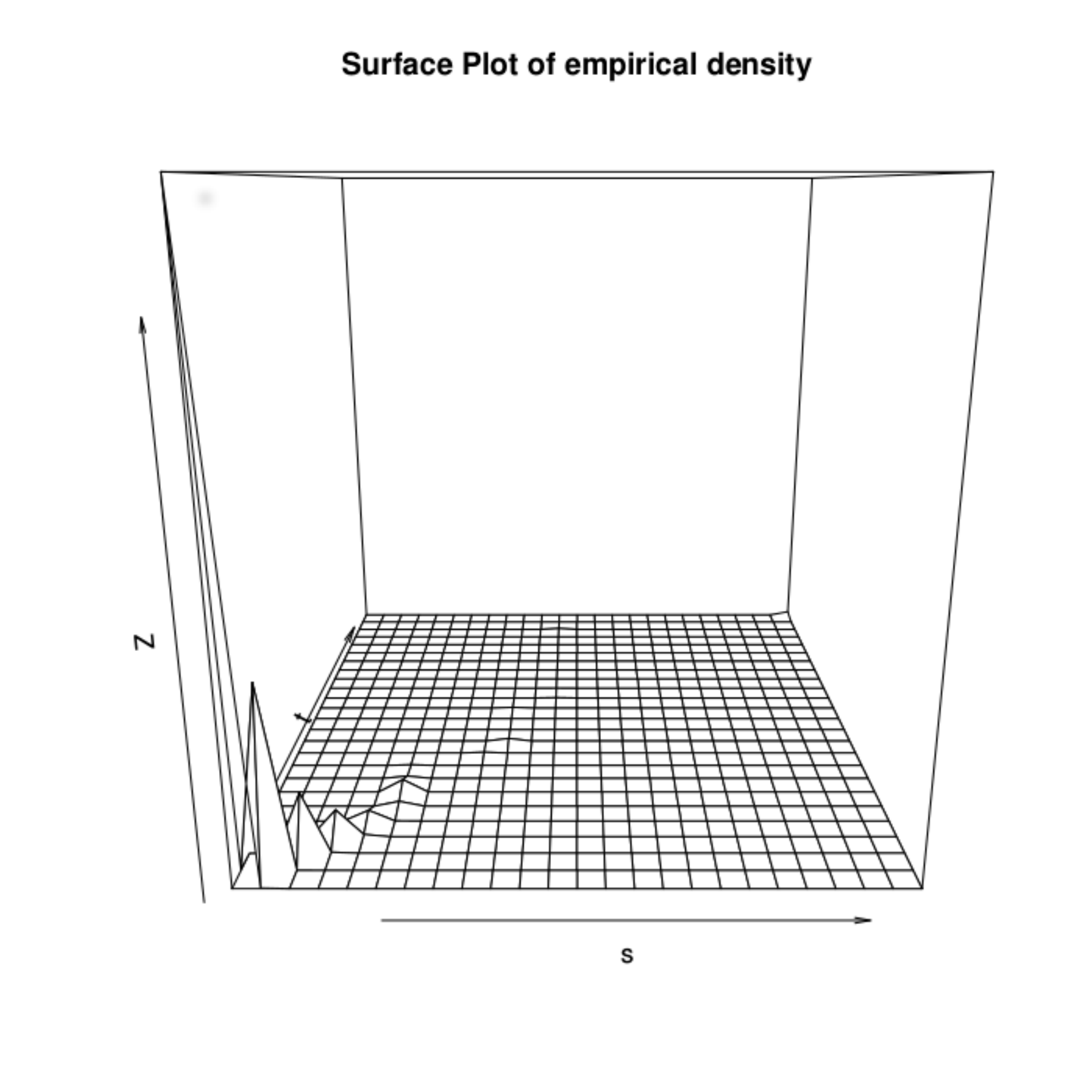} 
\caption{Two dimensional density plots of the transformed dataset \label{fig4}}
\end{center}
\end{figure}

  Empirical two dimensional density plot in Figure-\ref{fig4} verifies that Marshall-Olkin Pareto can be an alternative model for the transformed dataset.  Parameter estimates of this bivariate distribution based on sample size $n = 468$ are provided in Table-\ref{tab:DATA_AE1} and Table-\ref{tab:DATA_AE2} whereas parametric bootstrap confidence intervals are provided in Table-\ref{tab:DATA_Conf1} and Table-\ref{tab:Conf2} respectively.  All algorithms are used separately to calculate the estimates and confidence intervals. 


\begin{table}[!ht]
\begin{small}
\begin{center}
\begin{tabular}{|c|c|c|c|c|}\hline
 & parameters & $\mu_1$ & $\mu_2$ & $\sigma_1$ \\ 
Procedures &  &  &  & \\ \hline   
  Mod. 1 & &1 & 0.4 & 2.5594 \\   
  Mod. 2 (T) & &1 & 0.4 & 2.4936 \\
  Mod. 2 (WT) & &1 & 0.4 & 2.9459 \\  
  Mod. 3 & &1 & 0.4 & 3.2459 \\ 
  Mod. 4 & & 1 & 0.4 & 3.1802 \\ \hline  
\end{tabular}  
\caption{EM estimates of the parameters - Data Analysis}  
\label{tab:DATA_AE1}
\end{center}
\end{small}
\end{table}

\begin{table}[!ht]
\begin{small}
\begin{center}
\begin{tabular}{|c|c|c|c|c|}\hline
Procedures & $\sigma_2$ & $\alpha_0$ & $\alpha_1$ & $\alpha_2$ \\
 &  &  &  & \\ \hline   
  Mod. 1 & 0.7176 & 0.7557 & 0.8162 & 0.9792 \\   
  Mod. 2 (T) &  0.8164 & 0.759 & 0.7878 & 1.1267 \\
  Mod. 2 (WT) & 0.8162 & 0.7960 & 0.9209 & 1.0895  \\  
  Mod. 3 & 0.8163 & 0.8171 & 1.0100 & 1.0684\\ 
  Mod. 4 & 0.7947 & 0.8330 & 0.9667 & 1.0158 \\ \hline    
\end{tabular}  
\caption{EM estimates of the parameters - Data Analysis}  
\label{tab:DATA_AE2}
\end{center}
\end{small}
\end{table}

\begin{table}[!ht]
\begin{small}
\begin{center}
\begin{tabular}{|c|c|c|c|c|}\hline
 & parameters & $\mu_1$ & $\mu_2$ & $\sigma_1$ \\ 
Procedures &  &  &  & \\ \hline   
 Mod. 1 & & [1.00013, 1.0127] & [0.40002, 0.4031] & [1.5793, 2.9953]\\   
 Mod. 2 (T) & & [1.00013, 1.0127] & [0.40002, 0.4031] & [1.4952, 2.7412] \\
 Mod. 2 (WT) & & [1.0001, 1.0135] & [0.40002, 0.4032] & [1.5033, 3.9036] \\  
 Mod. 3 & & [1.00013, 1.0141] & [0.40002, 0.4032] & [2.3578, 4.9274] \\ 
 Mod. 4 & & [1.00013, 1.01397] & [0.40002, 0.4032] & [2.3343, 4.7869] \\  \hline  
\end{tabular}  
\caption{Confidence interval - Data Analysis}  
\label{tab:DATA_Conf1}
\end{center}
\end{small}
\end{table}

\begin{table}[!ht]
\begin{small}
\begin{center}
\begin{tabular}{|c|c|c|c|c|}\hline
 & $\sigma_2$ & $\alpha_0$ & $\alpha_1$ & $\alpha_2$ \\
Procedures &  &  &  & \\ \hline   
 Mod. 1 & [0.4554, 0.9018] & [0.5435, 0.8524] & [0.5378, 1.0407] & [0.6726, 1.3616]\\   
 Mod. 2 (T) & [0.4337, 1.1601] & [0.5234, 0.8712] & [0.4883, 0.9623] & [0.6829, 1.7491]\\
 Mod. 2 (WT) & [0.4448, 1.1833] & [0.5445, 0.9308] & [0.5090, 1.3183] & [0.6909, 1.7151] \\  
 Mod. 3 & [0.5731, 1.2017] & [0.5869, 0.9988] & [0.7788, 1.6505] & [0.7996, 1.7046]\\ 
 Mod. 4 & [0.5658, 1.1673] & [0.6089, 1.0025] & [0.7526, 1.5866] & [0.7770, 1.6220] \\  \hline  
\end{tabular}  
\caption{Confidence interval - Data Analysis}  
\label{tab:Conf2}
\end{center}
\end{small}
\end{table}

\section{Conclusion}

 We use different variations of EM algorithms for estimating the parameters of singular bivariate Pareto distribution.  All variations work quite well even for moderate sample size (say, 150, 250 etc).  Our algorithms outperforms the current state-of-art algorithm by Asimit et al. The data analysis makes the paper more interesting.  There was no discussion of any real-life data analysis in the previous paper.  We have also calculated parametric bootstrap confidence interval which is absent in Asimit et al.  The algorithm shown here can be applicable to other higher dimensional distributions with location and scale parameters. Given a real-life data set model selection can be an interesting issue among different other form of bivariate pareto distributions. More work is needed in this direction too.

\appendix

\section{}

\subsection{} \begin{eqnarray*} S(x_{1}, x_{2}) & = & P(X_{1} \geq x_{1}, X_{2} \geq x_{2}) = \begin{cases}  S_{1}(x_{1}, x_{2})  & \text{if $\frac{x_{1} - \mu_{1}}{\sigma_{1}} < \frac{x_{2} - \mu_{2}}{\sigma_{2}}$}\\
 S_{2}(x_{1}, x_{2})  & \text{if $\frac{x_{1} - \mu_{1}}{\sigma_{1}} > \frac{x_{2} - \mu_{2}}{\sigma_{2}}$}\\
 S_{0}(x) & \text{if $\frac{x_{1} - \mu_{1}}{\sigma_{1}} = \frac{x_{2} - \mu_{2}}{\sigma_{2}} = x $}
\end{cases}
\end{eqnarray*}

\begin{eqnarray*} S_{1}(x_{1}, x_{2}) = (1 + \frac{(x_{1} - \mu_{1})}{\sigma_{1}})^{-\alpha_{1}}(1 + \frac{x_{2} - \mu_{2}}{\sigma_{2}})^{-(\alpha_{0} + \alpha_{2})} \end{eqnarray*}
\begin{eqnarray*} S_{2}(x_{1}, x_{2}) = (1 + \frac{(x_{1} - \mu_{1})}{\sigma_{1}})^{-(\alpha_{0} + \alpha_{1})}(1 + \frac{x_{2} - \mu_{2}}{\sigma_{2}})^{\alpha_{2}} \end{eqnarray*}
\begin{eqnarray*} S_{0}(x) = (1 + x)^{-(\alpha_{0} + \alpha_{1} + \alpha_{2})} \end{eqnarray*} 

From the above expressions we can get the marginal of $X_{1}$ and $X_{2}$ taking $x_{1} \rightarrow \mu_1$ and $x_{2} \rightarrow \mu_2$,
\begin{eqnarray*} S_{X_{1}}(x_{1}) = \begin{cases} (1 + \frac{(x_{1} - \mu_{1})}{\sigma_{1}})^{-(\alpha_{0} + \alpha_{1})} & \text{if $x_{1} > \mu_{1}$}\\ 1 &  \text{otherwise} \end{cases} \end{eqnarray*}
\begin{eqnarray*} S_{X_{2}}(x_{2}) = \begin{cases} (1 + \frac{x_{2} - \mu_{2}}{\sigma_{2}})^{-(\alpha_{0} + \alpha_{2})} & \text{if $x_{1} > \mu_{1}$}\\ 1 & \text{otherwise} \end{cases} \end{eqnarray*}

\subsection{}  \begin{equation}
    P(\min\{ X_{1}, X_{2} \} \geq x) = P(X_{1} \geq x, X_{2} \geq x)
\end{equation}
This implies
\begin{equation}
    P(X_{1} > x, X_{2} > x) = P(U_{0} > x, U_{1} > x, U_{2} > x)\\ 
\end{equation}
Since, $$ X_{k} = \min\{ U_{0}, U_{k} \} $$
Therefore,
\begin{eqnarray*}
    P(X_{1} > x, X_{2} > x) & = &  P(U_{0} > x, U_{1} > x, U_{2} > x)\\ & = & (1 + \frac{x - \mu}{\sigma})^{-\alpha_{0}}(1 + \frac{x - \mu}{\sigma})^{-\alpha_{1}} (1 + \frac{x - \mu}{\sigma})^{-\alpha_{2}}\\ & = & (1 + \frac{x - \mu}{\sigma})^{-(\alpha_{0} + \alpha_{1} + \alpha_{2})}
\end{eqnarray*}

\subsection{}  It is easy to show that maximum likelihood estimate of location parameter is $\mu = \min\{ X_{1}, X_{2}, \cdots, X_{n} \}.$  Rest two equations can be obtained by plug-in the estimates of $\mu$ and taking derivative of Log-likelihood with respect to $\sigma$ and $\alpha$. Therefore the details of the proof is omitted.   

\nocite{AsimitVernicZitikis:2013}, \nocite{BalakrishnanGuptaKunduLeivaSanhueza:2011}, \nocite{BlockBasu:1974}, \nocite{CampbellAustin:2002}, \nocite{GuptaKundu:1999}, \nocite{Kundu:2012}, \nocite{KunduKumarGupta:2015}, \nocite{KhosraviKunduJamalizadeh:2015}, \nocite{MirhosseiniAminiKunduDolati:2015}, \nocite{McLachlanPeel:2004}, \nocite{Marshall:1996}, \nocite{RisticKundu:2015}, \nocite{RakonczaiZempleni:2012}, \nocite{SmithTawnColes:1997}, \nocite{Smith:1994}, \nocite{Tsay:2014}      

\bibliographystyle{chicago}

\bibliography{bvpa}

\begin{thebibliography}{}

\bibitem[\protect\citeauthoryear{Asimit, Furman, and Vernic}{Asimit
  et~al.}{2010}]{AsimitFurmanVernic:2010}
Asimit, A.~V., E.~Furman, and R.~Vernic (2010).
\newblock On a multivariate pareto distribution.
\newblock {\em Insurance: Mathematics and Economics\/}~{\em 46\/}(2), 308--316.

\bibitem[\protect\citeauthoryear{Asimit, Furman, and Vernic}{Asimit
  et~al.}{2016}]{AsimitFurmanVernic:2016}
Asimit, A.~V., E.~Furman, and R.~Vernic (2016).
\newblock Statistical inference for a new class of multivariate pareto
  distributions.
\newblock {\em Communications in Statistics-Simulation and Computation\/}~{\em
  45\/}(2), 456--471.

\bibitem[\protect\citeauthoryear{Asimit, Vernic, and Zitikis}{Asimit
  et~al.}{2013}]{AsimitVernicZitikis:2013}
Asimit, A.~V., R.~Vernic, and R.~Zitikis (2013).
\newblock Evaluating risk measures and capital allocations based on
  multi-losses driven by a heavy-tailed background risk: The multivariate
  pareto-ii model.
\newblock {\em Risks\/}~{\em 1\/}(1), 14--33.

\bibitem[\protect\citeauthoryear{Balakrishnan, Gupta, Kundu, Leiva, and
  Sanhueza}{Balakrishnan
  et~al.}{2011}]{BalakrishnanGuptaKunduLeivaSanhueza:2011}
Balakrishnan, N., R.~C. Gupta, D.~Kundu, V.~Leiva, and A.~Sanhueza (2011).
\newblock On some mixture models based on the birnbaum--saunders distribution
  and associated inference.
\newblock {\em Journal of Statistical Planning and Inference\/}~{\em 141\/}(7),
  2175--2190.

\bibitem[\protect\citeauthoryear{Block and Basu}{Block and
  Basu}{1974}]{BlockBasu:1974}
Block, H.~W. and A.~Basu (1974).
\newblock A continuous bivariate exponential extension.
\newblock {\em Journal of the American Statistical Association\/}~{\em
  69\/}(348), 1031--1037.

\bibitem[\protect\citeauthoryear{Campbell and Austin}{Campbell and
  Austin}{2002}]{CampbellAustin:2002}
Campbell, J.~I. and S.~Austin (2002).
\newblock Effects of response time deadlines on adults’ strategy choices for
  simple addition.
\newblock {\em Memory \& Cognition\/}~{\em 30\/}(6), 988--994.

\bibitem[\protect\citeauthoryear{D}{D}{2012}]{Kundu:2012}
D, K. (2012).
\newblock On sarhan-balakrishnan bivariate distribution.
\newblock {\em Journal of Statistics Applications \& Probability\/}~{\em
  1\/}(3), 163.

\bibitem[\protect\citeauthoryear{Dey and Kundu}{Dey and
  Kundu}{2012}]{DeyKundu:2012}
Dey, A.~K. and D.~Kundu (2012).
\newblock Discriminating between bivariate generalized exponential and
  bivariate weibull distributions.
\newblock {\em Chilean Journal of Statistics\/}~{\em 3\/}(1), 93--110.

\bibitem[\protect\citeauthoryear{Falk and Guillou}{Falk and
  Guillou}{2008}]{FalkGuillou:2008}
Falk, M. and A.~Guillou (2008).
\newblock Peaks-over-threshold stability of multivariate generalized pareto
  distributions.
\newblock {\em Journal of Multivariate Analysis\/}~{\em 99\/}(4), 715--734.

\bibitem[\protect\citeauthoryear{fbvpot~by Chris~Ferro}{fbvpot~by
  Chris~Ferro}{2015}]{Stephenson:2015}
fbvpot~by Chris~Ferro, A. S.~F. (2015).
\newblock {\em evd : Functions for extreme value distributions}.

\bibitem[\protect\citeauthoryear{Gupta and Kundu}{Gupta and
  Kundu}{1999}]{GuptaKundu:1999}
Gupta, R.~D. and D.~Kundu (1999).
\newblock Theory \& methods: Generalized exponential distributions.
\newblock {\em Australian \& New Zealand Journal of Statistics\/}~{\em
  41\/}(2), 173--188.

\bibitem[\protect\citeauthoryear{Hanagal}{Hanagal}{1996}]{Hanagal:1996}
Hanagal, D.~D. (1996).
\newblock A multivariate pareto distribution.
\newblock {\em Communications in Statistics-Theory and Methods\/}~{\em
  25\/}(7), 1471--1488.

\bibitem[\protect\citeauthoryear{Khosravi, Kundu, and Jamalizadeh}{Khosravi
  et~al.}{2015}]{KhosraviKunduJamalizadeh:2015}
Khosravi, M., D.~Kundu, and A.~Jamalizadeh (2015).
\newblock On bivariate and a mixture of bivariate birnbaum--saunders
  distributions.
\newblock {\em Statistical Methodology\/}~{\em 23}, 1--17.

\bibitem[\protect\citeauthoryear{Kundu and Dey}{Kundu and
  Dey}{2009}]{KunduDey:2009}
Kundu, D. and A.~K. Dey (2009).
\newblock Estimating the parameters of the marshall--olkin bivariate weibull
  distribution by em algorithm.
\newblock {\em Computational Statistics \& Data Analysis\/}~{\em 53\/}(4),
  956--965.

\bibitem[\protect\citeauthoryear{Kundu and Gupta}{Kundu and
  Gupta}{2009}]{KunduGupta:2009}
Kundu, D. and R.~D. Gupta (2009).
\newblock Bivariate generalized exponential distribution.
\newblock {\em Journal of Multivariate Analysis\/}~{\em 100\/}(4), 581--593.

\bibitem[\protect\citeauthoryear{Kundu and Gupta}{Kundu and
  Gupta}{2010}]{KunduGupta:2010}
Kundu, D. and R.~D. Gupta (2010).
\newblock A class of absolutely continuous bivariate distributions.
\newblock {\em Statistical Methodology\/}~{\em 7\/}(4), 464--477.

\bibitem[\protect\citeauthoryear{Kundu, Kumar, and Gupta}{Kundu
  et~al.}{2015}]{KunduKumarGupta:2015}
Kundu, D., A.~Kumar, and A.~K. Gupta (2015).
\newblock Absolute continuous multivariate generalized exponential
  distribution.
\newblock {\em Sankhya B\/}~{\em 77\/}(2), 175--206.

\bibitem[\protect\citeauthoryear{Marshall}{Marshall}{1996}]{Marshall:1996}
Marshall, A.~W. (1996).
\newblock Copulas, marginals, and joint distributions.
\newblock {\em Lecture Notes-Monograph Series\/}, 213--222.

\bibitem[\protect\citeauthoryear{Marshall and Olkin}{Marshall and
  Olkin}{1967}]{MarshallOlkin:1967}
Marshall, A.~W. and I.~Olkin (1967).
\newblock A multivariate exponential distribution.
\newblock {\em Journal of the American Statistical Association\/}~{\em
  62\/}(317), 30--44.

\bibitem[\protect\citeauthoryear{McLachlan and Peel}{McLachlan and
  Peel}{2004}]{McLachlanPeel:2004}
McLachlan, G. and D.~Peel (2004).
\newblock {\em Finite mixture models}.
\newblock John Wiley \& Sons.

\bibitem[\protect\citeauthoryear{Mirhosseini, Amini, Kundu, and
  Dolati}{Mirhosseini et~al.}{2015}]{MirhosseiniAminiKunduDolati:2015}
Mirhosseini, S.~M., M.~Amini, D.~Kundu, and A.~Dolati (2015).
\newblock On a new absolutely continuous bivariate generalized exponential
  distribution.
\newblock {\em Statistical Methods \& Applications\/}~{\em 24\/}(1), 61--83.

\bibitem[\protect\citeauthoryear{Nelsen}{Nelsen}{2007}]{Nelsen:2007}
Nelsen, R.~B. (2007).
\newblock {\em An introduction to copulas}.
\newblock Springer Science \& Business Media.

\bibitem[\protect\citeauthoryear{Rakonczai and Zempl{\'e}ni}{Rakonczai and
  Zempl{\'e}ni}{2012}]{RakonczaiZempleni:2012}
Rakonczai, P. and A.~Zempl{\'e}ni (2012).
\newblock Bivariate generalized pareto distribution in practice: models and
  estimation.
\newblock {\em Environmetrics\/}~{\em 23\/}(3), 219--227.

\bibitem[\protect\citeauthoryear{Risti{\'c} and Kundu}{Risti{\'c} and
  Kundu}{2015}]{RisticKundu:2015}
Risti{\'c}, M.~M. and D.~Kundu (2015).
\newblock Marshall-olkin generalized exponential distribution.
\newblock {\em Metron\/}~{\em 73\/}(3), 317--333.

\bibitem[\protect\citeauthoryear{Sarhan and Balakrishnan}{Sarhan and
  Balakrishnan}{2007}]{SarhanBalakrishnan:2007}
Sarhan, A.~M. and N.~Balakrishnan (2007).
\newblock A new class of bivariate distributions and its mixture.
\newblock {\em Journal of Multivariate Analysis\/}~{\em 98\/}(7), 1508--1527.

\bibitem[\protect\citeauthoryear{Smith}{Smith}{1994}]{Smith:1994}
Smith, R.~L. (1994).
\newblock Multivariate threshold methods.
\newblock In {\em Extreme Value Theory and Applications}, pp.\  225--248.
  Springer.

\bibitem[\protect\citeauthoryear{Smith, Tawn, and Coles}{Smith
  et~al.}{1997}]{SmithTawnColes:1997}
Smith, R.~L., J.~A. Tawn, and S.~G. Coles (1997).
\newblock evd : Functions for extreme value distributions.
\newblock {\em https://cran.r-project.org\/}~{\em 84\/}(2), 249--268.

\bibitem[\protect\citeauthoryear{Tsay}{Tsay}{2014}]{Tsay:2014}
Tsay, R.~S. (2014).
\newblock {\em An introduction to analysis of financial data with R}.
\newblock John Wiley \& Sons.

\bibitem[\protect\citeauthoryear{Yeh}{Yeh}{2000}]{Yeh:2000}
Yeh, H.~C. (2000).
\newblock Two multivariate pareto distributions and their related inferences.
\newblock {\em Bulletin of the Institute of Mathematics, Academia
  Sinica.\/}~{\em 28\/}(2), 71--86.

\bibitem[\protect\citeauthoryear{Yeh}{Yeh}{2004}]{Yeh:2004}
Yeh, H.-C. (2004).
\newblock Some properties and characterizations for generalized multivariate
  pareto distributions.
\newblock {\em Journal of Multivariate Analysis\/}~{\em 88\/}(1), 47--60.

\end{thebibliography}

\end{document}